\documentclass[%
 aip,
 amsmath,amssymb,
 reprint,%
]{revtex4-2}

\usepackage{graphicx}% Include figure files
\usepackage{dcolumn}% Align table columns on decimal point
\usepackage{bm}% bold math
\usepackage[utf8]{inputenc}
\usepackage[T1]{fontenc}
\usepackage{mathptmx}
\usepackage{etoolbox}

\usepackage{xcolor}

\makeatletter
\def\@email#1#2{%
 \endgroup
 \patchcmd{\titleblock@produce}
  {\frontmatter@RRAPformat}
  {\frontmatter@RRAPformat{\produce@RRAP{*#1\href{mailto:#2}{#2}}}\frontmatter@RRAPformat}
  {}{}
}%
\makeatother
\begin{document}

\preprint{AIP/123-QED}

\title[]{Molecular-sized bubbles in a liquid: free energy of formation beyond the capillarity approximation}
% Force line breaks with \\
\author{Jo\"el Puibasset}
 \email{puibasset@cnrs-orleans.fr}
\affiliation{ 
ICMN, CNRS, Universit\'e d'Orl\'eans, 1b Rue de la F\'erollerie, CS 40059, 45071 Orl\'eans Cedex 02, France }

\date{\today}% It is always \today, today,

\begin{abstract}
We investigate the transient bubbles that spontaneously appear in a simple liquid using molecular simulations. The objective is to deduce the free-energy of formation of the bubbles $W(s)$ from the bubble size distribution $p(s)$ through the hypothesis of a Boltzmann distribution: $W(s) = -kT \ln p(s)$. The bubbles are detected and characterized using a method based on a grid superimposed on the liquid, efficient for bubbles larger than the grid mesh. We first investigate how the results are affected by the mesh choice, and show that using several mesh values allows to detect bubbles in a wide range of sizes with minimal computing cost. The free-energy of formation of a bubble can then be deduced for a large range of sizes, with particular emphasis in the region of vanishing bubbles scarcely investigated in previous works. We first show that the usual Boltzmann relation has to be modified when the bubble size is characterized by its volume. In particular, the bubble volume distribution diverges for a vanishing bubble, which should be taken into account before calculating its free-energy of formation from the above formula. An analytical expansion, valid for any interacting spherical molecules, confirms this observation. We then show that the capillarity approximation fails for small bubbles: an extra contribution, linear with the bubble radius, has to be added to the usual quadratic (surface) and cubic (volume) contributions to the free-energy. This extra term most probably relates to the irregular shape of the tiny bubbles.
\end{abstract}

\maketitle

\section{Introduction}
Among density fluctuations appearing spontaneously in a liquid, the voids or cavities (empty spaces between molecules) and bubbles (region filled with vapor) play an important role regarding its dynamic or thermodynamic properties, for instance its ability to solubilize molecules.\cite{RN2963, RN2859, RN2875, RN2984, RN2869, RN2961, RN3180, RN2861, RN2871, RN2983}

These fluctuations also play a crucial role in phase transitions, in particular during the liquid-to-vapor transition of a superheated or stretched liquid.\cite{RN2997, RN2760, RN2758, RN2715, RN2845, RN2972, RN2996, RN2991, RN3000, RN2989} In this context, the popular Classical Nucleation Theory (CNT) relies on the idea that bubbles can grow from seeds formed by spontaneous cavities.\cite{RN548, RN512, RN2920} Quantitatively, bubbles can reach the size $s$ with a probability $p$ proportional to the Boltzmann factor $\exp [ -W(\{ s \})/kT ] $, where $W(\{ s \})$ is the free energy of formation of the bubble, $k$ is Boltzmann's constant, $T$ is the temperature, and $\{ s \}$ is a set of variables that characterizes the bubble size and shape (volume, gyration radius, aspect ratio etc.).\cite{RN2076, RN2859, RN2913} To simplify the notations, the braces will be dropped, and an expression like ``bubble size $s$'', used in a general context, should be understood as including a description of the shape of the bubble. The main objective of this work is to assess the validity of such a relationship between $p$ and $W$ in the early stages of bubble formation. In particular, one has to determine the prefactor that fixes the origin of the free energy, and check that it is well behaved for a vanishing bubble.

A priori, the bubble can adopt any shape, possibly including strong distortions.\cite{RN2756, RN2801, RN2968, RN2985, RN2770, RN598, RN2903, RN2766, RN2797} However, it is generally admitted that, at least for bubbles large enough, the surface energy cost will favor spherical shapes. In CNT, the shape is assumed to be spherical for any size, which is parametrized by its radius $r$. In the capillarity approximation, the free energy of formation contains surface and volume contributions:
\begin{equation}
 W(r) = 4\pi r^2 \gamma + \frac{4}{3} \pi r^3 \Delta  \label{eq_w_cnt},
\end{equation}
where $\gamma$ is the surface tension and $\Delta$ is for instance a pressure difference in the case of bubble nucleation in a liquid at constant pressure and temperature. While the volume contribution $\Delta$ changes sign when crossing the coexistence line, the surface contribution, which dominates for tiny bubbles, is always positive and generates an energetic cost. The equation then tells us that the first stage of bubble growth is always disfavored, and that tiny bubbles have similar probabilities to occur spontaneously on both sides of the equilibrium line, i.e. in the stable as well as in the metastable liquid, with a probability approaching one for vanishing cavities.\cite{RN3177, RN3141} 

However, nothing can be said about the \emph{number} of spontaneous bubbles of size $s$ in the liquid, although, in many cases, one is rather interested in this quantity: for instance in CNT for the calculation of the nucleation rate, or in molecular simulations where the number of bubbles is directly measurable. In the context of condensation in a supersaturated vapor or crystallization in a supercooled liquid, the number $N_n$ of nuclei of size $s=n$, where $n$ is defined as its number of particles, has been shown, when clusters are rare, to verify\cite{RN3141, RN3145, RN1956, RN3140, RN2932, RN3139, RN2913, RN2911}
\begin{equation}
N_n = N e^{-W(n)/kT}  \label{eq_boltz_Nn}
\end{equation}
where $N$ is the total number of molecules in the system and $W(n)$ is the free energy of formation of a nucleus of size $n$. Note that $N_n$ is extensive, while the corresponding $N_n/N$ has been referred to as an ``intensive'' probability.\cite{RN1956, RN3140} In the case of bubble formation in a liquid, their size is a continuum, and one defines the density of bubbles of size $s$ as $p(s) = dN_s/ds$ where $dN_s$ is the infinitesimal number of bubbles of size between $s$ and $s+ds$. \cite{RN3172} $p(s)$ is again extensive, and one expects that an equation similar to Eq~\ref{eq_boltz_Nn} should hold for $p(s)$:
\begin{equation}
p(s) = Q e^{-W(s)/kT} . \label{eq_boltz_ps}
\end{equation}
It is however unclear how to generalize the arguments developed in the context of discrete clusters (Eq~\ref{eq_boltz_Nn}) to the case of continuous ones (Eq~\ref{eq_boltz_ps}).\cite{RN2913, RN1956} As previously, an intensive bubble size density can be defined. For instance, several authors introduced a normalization by the volume $V$ of the system: $\rho(s) = p(s)/V$.\cite{RN2801, RN3172, RN2966} To our knowledge, $Q$, or equivalently $\rho_0 = Q/V$, has generally been taken as a constant, and adjusted so that $W(0)=0$.\cite{RN2801} In the case of stretched water, Menzl et al.\cite{RN2801} found $\rho_0 = Q/V = 0.022$ nm$^{-6}$. 
It is however unclear how to define the free energy of a bubble of size exactly zero ($s=0$), and, therefore, it is generally assumed to be the limiting value of $w(s)$ for a vanishing bubble ($s \rightarrow 0)$. This however supposes that the corresponding limit does exist.

\begin{figure}[t]
\includegraphics[width=0.9\columnwidth]{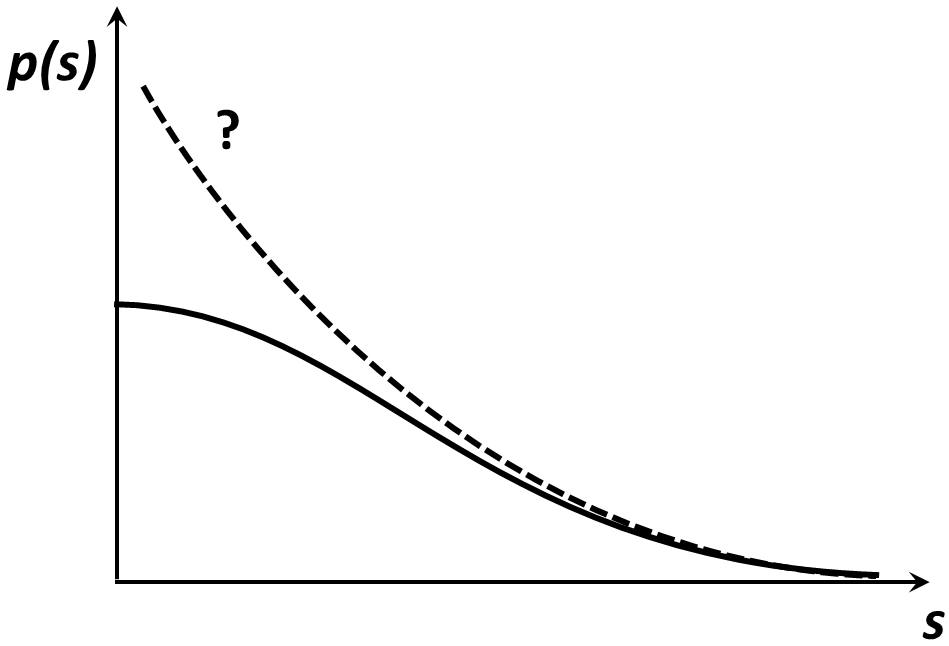}
\caption{\label{fig_ps} Schematic representation of $p(s)$, the density of bubbles of size $s$ in a liquid, as given by Eqs~\ref{eq_w_cnt} and \ref{eq_boltz_ps}. If $Q$ is constant or varies smoothly with $s$, $p(0)$ is finite (solid line). However, our simulations suggest that $p(s)$ may diverge around $s=0$ when the bubble size is characterized by its volume (dotted line).}
\end{figure}

Figure~\ref{fig_ps} gives a schematic representation of $p(s)$ as given by Eq~\ref{eq_boltz_ps} with the hypothesis that $Q$ is constant or varies smoothly around $s=0$ (solid line).
However, the simulation results presented in this work suggest that $Q$ may not be constant, and may even present, in some cases, a divergence at vanishing size (see Fig.~\ref{fig_ps}, dotted line). Note however that, if the condition that bubbles do not overlap is satisfied, the total volume occupied by the bubbles $\int_0^\infty vp(v)dv$ has to be finite ($\leq V$), which puts a limit on the degree of divergence of $Q$. Integrability of the distribution imposes an even stronger limit.  

The last point to be mentioned is that the exact expression for $p(s)$ is expected to explicitly depend on the chosen parameter used to characterize the bubble size, even when there is a one to one correspondence between them. For example, a spherical bubble may either be characterized by its radius $r$ or its volume $v$. Since the number of bubbles between $s$ and $s+ds$ is the same for $s=r$ or $v$, one has $dN_s = p_r(r)dr = p_v(v)dv$ where we have introduced an index to differentiate the two mathematical expressions for $p$. Therefore, 
\begin{equation}
p_r(r) = 4 \pi r^2 p_v \left( \frac{4}{3} \pi r^3 \right) ,  \label{eq_pr_pv}
\end{equation}
and, necessarily, in Eq~\ref{eq_boltz_ps}, $Q$ cannot be a constant in both representations. 
In literature, in most cases, one uses volume distributions to calculate free energies. This is the case in the context of condensation of vapors or crystallization,\cite{RN2901, RN2932, RN3139, RN3140, RN1956, RN3145, RN2911, RN2915, RN598, RN2905, RN2903, RN2907, RN2923, RN2899} where the natural definition of the nucleus size is its number of atoms, but also in the context of bubble nucleation.\cite{RN2797, RN2847, RN2801} On the other hand, the radial distribution is more scarcely used.\cite{RN2871}
Two questions then arise. Is it possible to determine $Q(s)$ for a given choice of size parameter? And, is there a particular choice for which $Q$ would be constant?

Experiments cannot help because the tiny spontaneous bubbles in the liquid are unobservable, but molecular simulations are reliable enough at this scale to provide an efficient alternative. The principle consists in performing a Molecular Dynamics or a Monte Carlo run of a stable or metastable liquid, and following the spontaneous growth and shrink of bubbles. Defining a bubble when this one is large is not difficult. However, when its size approaches atomic scale, it is more involved. There are different methods, all relying on an estimate of the local fluid density, but using quite different approaches, and giving a priori different results for small bubbles.\cite{RN2968, RN2797, RN1184, RN2847} In this paper, we focus on a popular method proposed by Wang et al.,\cite{RN1184} denoted as the W-method in the following. The principle consists in analyzing the liquid structure based on a geometric criterion and a grid superimposed on the molecular configuration. For a large grid mesh, the method is fast, but it is unable to capture accurately the shape of the bubbles, and completely misses the smaller ones. Conversely, the smaller the grid mesh, the more precise the characterization of the bubble shape and volume. But the computational cost increases prohibitively, in particular for large systems. In practice, the mesh is generally taken slightly smaller than the molecular size.

The objective of this work is to use the W-method to characterize the bubble size distribution $p(s)$, with a particular focus for $s \rightarrow 0$. The molecular model is presented, as well as a brief reminder of the W-method. A careful analysis of the influence of the grid mesh is presented, in particular for small bubbles. It is shown that the bubble size distribution converges toward a definite distribution when the grid mesh is chosen arbitrarily small, which allows to extrapolate $p(s)$ to vanishing bubble sizes. An unexpected divergence is observed at $s=0$ when the bubble size is characterized by its volume ($s=v$), which has been confirmed by an analytical calculation that fits very well with the numerical results for tiny bubbles. This divergence however disappears if the bubble size is characterized by its radius, allowing to define a free energy of formation. It is then shown that the capillarity approximation fails to reproduce the free energy of the smallest bubbles.

\section{Numerical details}
\subsection{Molecular model} 
The ideas developed in this study are generic and depend essentially on the atomistic structure of the liquid and the method used to define the cavities or bubbles. We therefore focus on a system of particles interacting through the (12-6) Lennard-Jones potential, depending on the two constants $\epsilon$ and $\sigma$ corresponding to the depth of the potential well and the molecular kinetic diameter respectively. The potential is truncated and quadratically shifted at a distance taken equal to 3$\sigma$.\cite{RN154,RN51,RN1330} All quantities will be expressed in reduced units, using $\sigma$ for distances, $\epsilon$ for energies, $\epsilon/k$ for temperatures, and $\epsilon/\sigma^3$ for pressures. This model is expected to apply to simple liquids with spherical molecules. For more complex molecules, highly anisotropic or flexible, more refined definitions of bubbles or cavities would be required, with probably different conclusions. For instance, in the case of network-forming liquids like water, the grid method has to be adapted.\cite{RN2847}

The property we are interested in, i.e. the spontaneous formation of cavities, is mainly controlled by thermal activation. We therefore perform a Monte Carlo simulation, but molecular dynamics would give identical results. To enhance statistics and favor density fluctuations, we work at the reduced temperature $kT/\epsilon=1.0$ and at a pressure in the metastable region slightly below the saturating pressure. However, since the method we use to detect the cavities relies on a mapping of the liquid with a grid, because of the periodic boundary conditions, the grid mesh has to follow the fluctuations of the simulation box during an isobaric run. Since we precisely want to study the impact of a particular choice for the grid mesh, we prefer to avoid mesh fluctuations and work at constant box volume in the Grand Canonical ensemble and impose the chemical potential instead of the pressure. More specifically, we work at $\ln \left( z \sigma^3 \right) = -3.20$ where $z = e^{\mu /kT}/\Lambda^3 $ is the activity, $\mu$ is the chemical potential and $\Lambda$ is the thermal de Broglie wavelength.\cite{RN154, RN51} Note that the activity is the natural parameter entering the Grand Canonical Monte Carlo, while the chemical potential would depend explicitly on the mass of the molecules. In these conditions, the average reduced pressure is $p\sigma^3/\epsilon=0.026$, a value below the reduced saturating pressure $p_{\rm sat}\sigma^3/\epsilon=0.055$ at the working temperature, which places the system in a slightly metastable state (superheated). Although the formation of tiny bubbles is slightly favored by the metastability, no liquid-to-vapor transition was observed during the simulation runs: the metastability is too weak to allow the system to overcome the energy barrier. It is also emphasized that we have not used any bias to sample the system (except to acquire information for large bubbles, see below): the observed bubbles are all spontaneous. A cubic system of edge $L = 18\sigma$ has been considered, with periodic boundary conditions, containing on average 3300 particles.

\subsection{Bubble identification and characterization}
\begin{figure}[t]
\includegraphics[width=0.9\columnwidth]{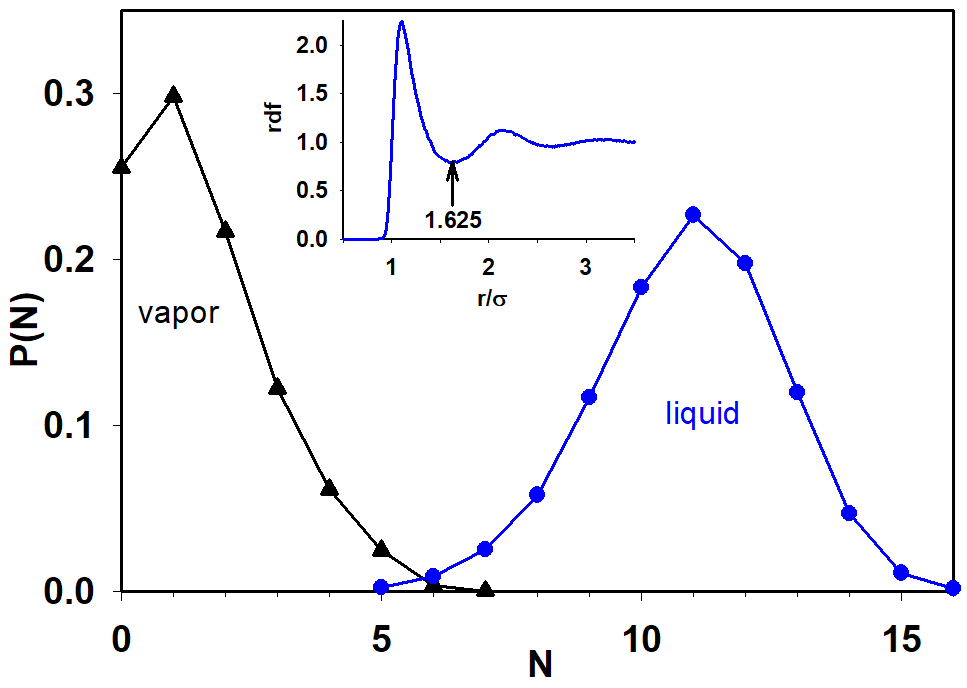}
\caption{\label{fig_P_N} Distributions in the vapor and liquid phases of the number of neighbors within the Stillinger sphere defined by the first minimum of the radial distribution function (insert) of the liquid in our thermodynamic conditions ($kT/\epsilon = 1$ and $\ln \left( z \sigma^3 \right) = -3.20$)}
\end{figure}

The bubbles appearing in the liquid are identified and characterized by estimating the local fluid density. Is this work, we focus on a method which consists in working on a discrete grid, and evaluate the impact of a particular choice for the grid mesh. We anticipate that using different mesh values will produce different results for the smallest fluid structures, but the largest ones should be unchanged. In this study we have chosen a four-steps algorithm which has proven to be very efficient, developed by Wang et al.\cite{RN1184} and denoted as W-method. We proceed as follows:

(i) The first step consists in identifying liquid-like and vapor-like molecules in the system. For each molecule, one counts the number of neighbors closer than $1.625\sigma$, corresponding to the first minimum of the radial distribution function in our thermodynamic conditions (Stillinger's criterion).\cite{RN3144} The corresponding distribution of the number of neighbors exhibits two well separated lobes (below 5 and above 6 neighbors, see Fig.~\ref{fig_P_N}).
Molecules having five or less neighbors are labeled as vapor-like, while those having six or more neighbors are labeled as liquid-like.\cite{RN1956} In the next steps, the vapor-like molecules are discarded, and only liquid-like ones are considered.

\begin{figure}[t]
\includegraphics[width=0.8\columnwidth]{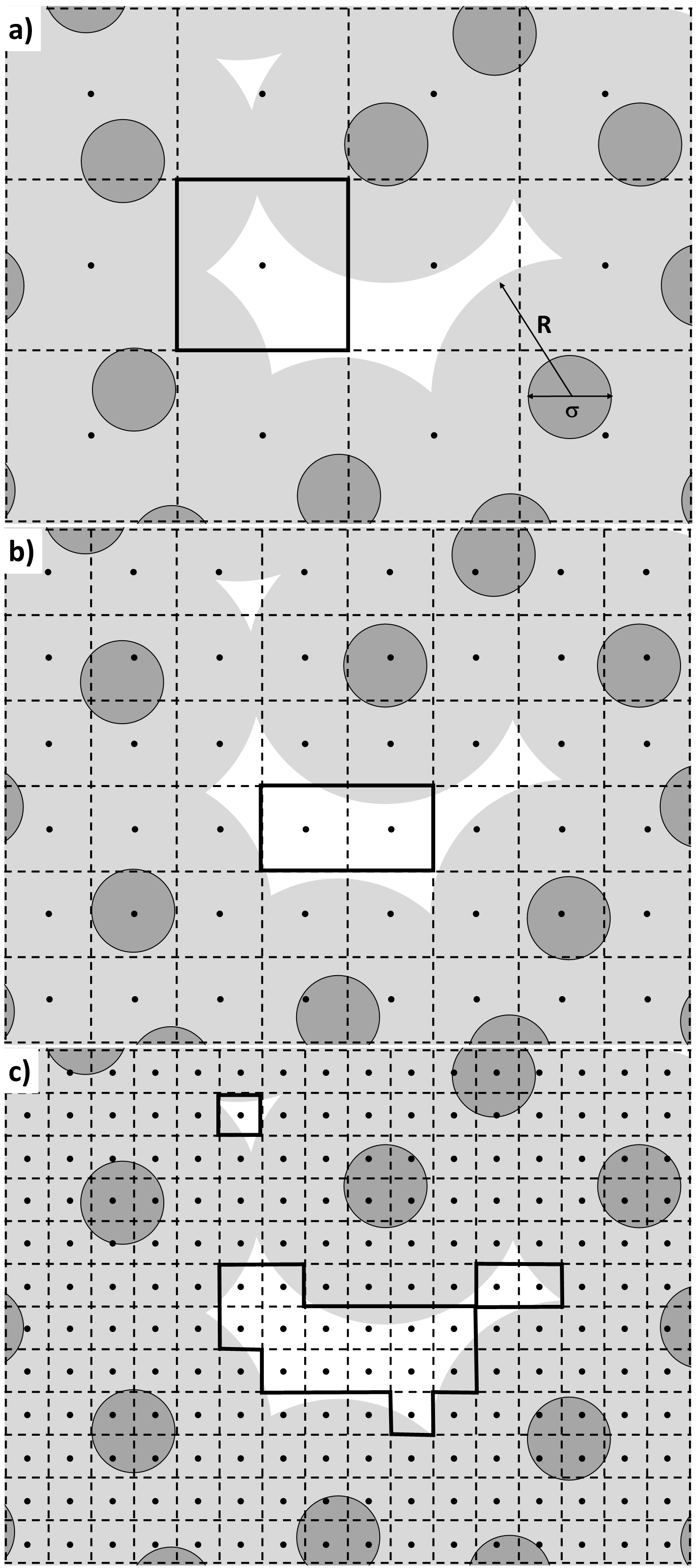}
\caption{\label{fig_grid} Illustration of the W-method\cite{RN1184} in the two-dimensional case. The dark gray disks represent the liquid-like molecules (kinetic diameter $\sigma$), and the light gray corona are their Stillinger\cite{RN3144} disks (radius $R$ taken equal to $1.625\sigma$ like in the three-dimensional case). Panels (a), (b) and (c) correspond to three grids with meshes $l_{\rm cell} = 2\sigma$, $\sigma$ and $0.5\sigma$ respectively. Cells whose centers (black dots) do not fall within any Stillinger sphere are marked as vapor. Bubbles are defined as clusters of vapor cells (highlighted regions with a thick boundary). For a vanishing mesh, this definition of bubbles converges toward the white regions between Stillinger disks. }
\end{figure}

(ii) The simulation box is then divided into cubic cells of edge $l_{\rm cell}$ (see Fig.~\ref{fig_grid} for an illustration in two dimensions).
In practice, $l_{\rm cell}$ is chosen smaller that the particle size to gain insight in the fluid structure at the molecular level, but not too small to keep the procedure computationally fast. A good compromise commonly accepted is $l_{\rm cell}=0.5\sigma$.\cite{RN1184} In this work, we explicitly vary $l_{\rm cell}$ to evaluate the influence of this parameter, starting with a significantly larger value $l_{\rm cell} = 2\sigma$, and reducing it by factors of two down to $2^{-4}\sigma$. The volume of the cell is denoted as $\delta = l_{\rm cell}^3$, ranging from $2^{-12}\sigma^3$ to $2^3\sigma^3$.

(iii) The grid cells in the vicinity of a liquid-like molecule (center-to-center distance $<1.625\sigma$) are marked as liquid, while the others are marked as vapor (see Fig.~\ref{fig_grid}).

(iv) We then perform a cluster analysis on the vapor cells to define the bubbles: two cells are considered to belong to the same bubble if they are connected through the faces, edges or vertices, i.e. their center-to-center distance is less or equal to $\sqrt{3} l_{\rm cell}$ (In Fig.~\ref{fig_grid}, the corresponding criterion in two dimensions is $\sqrt{2} l_{\rm cell}$). The size of the bubble may either be defined as the number $n$ of cells in the cluster or its total volume $v=n \delta$.  

Step (i) is the most important since it defines the fluid structure in terms of liquid-like and vapor-like molecules. Steps (ii) to (iv) correspond to the fast discrete numerical procedure used to characterize the bubbles. In the vanishing limit $l_{\rm cell} \rightarrow 0$, the discrete procedure (ii) to (iv) gives the exact volume and shape of the bubbles defined as one-piece regions of space out of the Stillinger spheres associated to the liquid-like molecules in step (i). Note that these regions can either contain vapor-like molecules, as expected for a bubble, or be completely empty, in which case it would be more appropriate to call them cavities or voids. However, we will not do the distinction in the following. 

\subsection{Large bubbles and umbrella sampling}
The bubbles forming spontaneously in a liquid are in general quite small. The probability of occurrence of larger bubbles decreasing very fast, a biased method is required to study their statistics. We have used the standard umbrella sampling scheme of Torrie and Valleau\cite{RN2885} to calculate the probability distribution $p(r)$ for $r \geq \sigma$. This method has been a breakthrough for exploring the nucleation barrier of various systems.\cite{RN1956, RN2911, RN2915, RN2905, RN2893, RN2895, RN2907, RN2903, RN2847, RN2801, RN2936, RN2909} We have also taken advantage of a recently developed method \cite{RN2954, RN3149} that cures approximations pointed out by Goswami and coworkers.\cite{RN2899} Although the method applies in principle to any size, we are limited in practice by the shortest dimension of the simulation box (18 $\sigma$ in our case) and the periodic boundary conditions: bubbles larger than $r = 4 \sigma$ have a significant probability of interacting with their own images, or of transforming irreversibly into cylindrical bubbles crossing the box. Using larger boxes being out of the scope of this work, the distribution will be limited to $r \leq 4 \sigma$.

\section{Results}
\subsection{Molecular configuration analysis}
\begin{figure}[t]
\includegraphics[width=0.9\columnwidth]{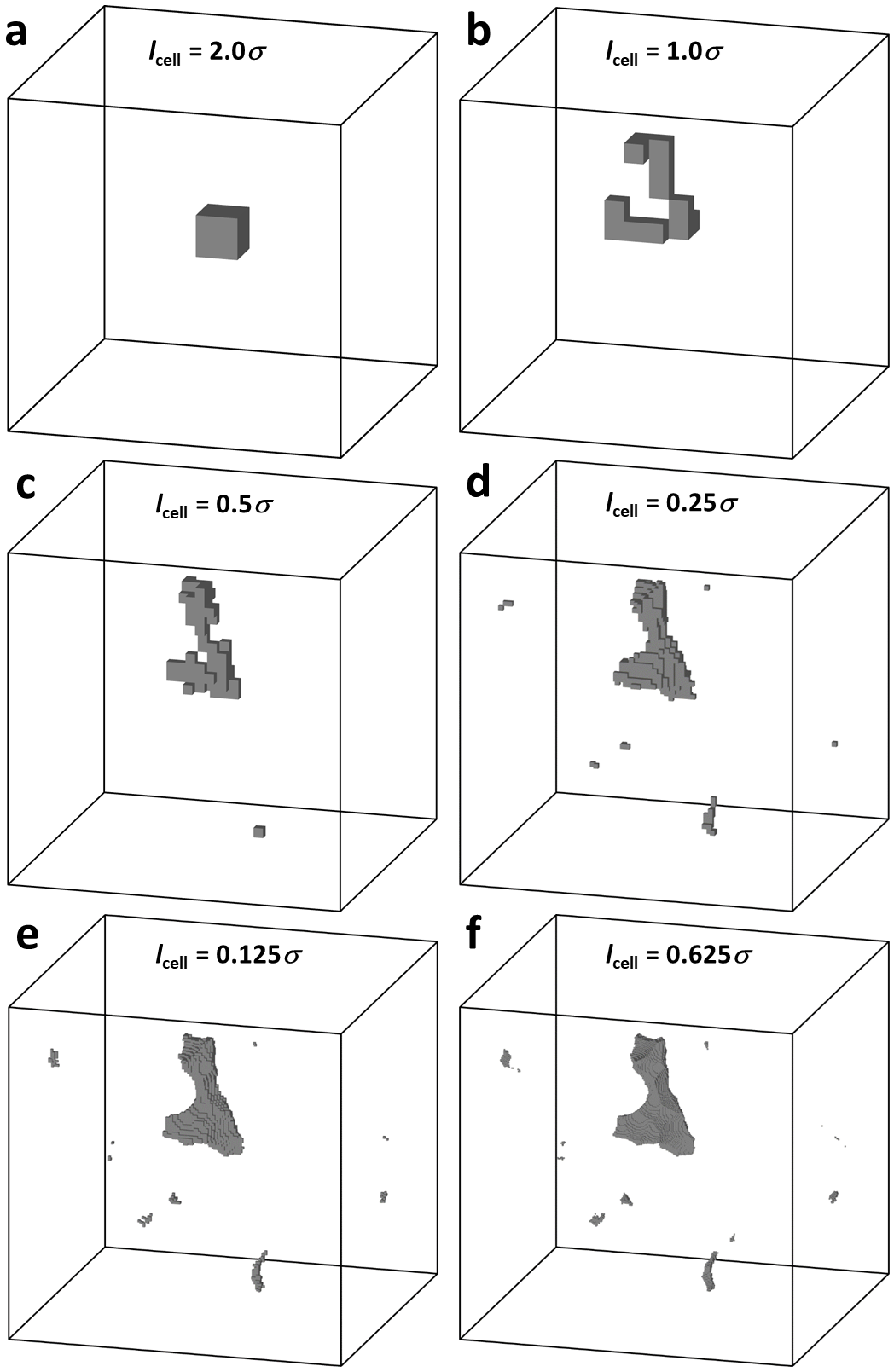}
\caption{\label{fig_snapshots_17} Schematic representation rendered with Ovito software\cite{RN3044} of the bubbles detected with the W-method for a particular molecular configuration. The grayed cells correspond to the vapor-like cells obtained in step (iii) of the method. Panels (a) to (f) correspond to the \emph{same} molecular configuration analyzed with a grid mesh $l_{\rm cell}$ decreasing from 2$\sigma$ to $2^{-4}\sigma$ by factors of two.}
\end{figure}

\begin{figure}[]
\includegraphics[width=0.9\columnwidth]{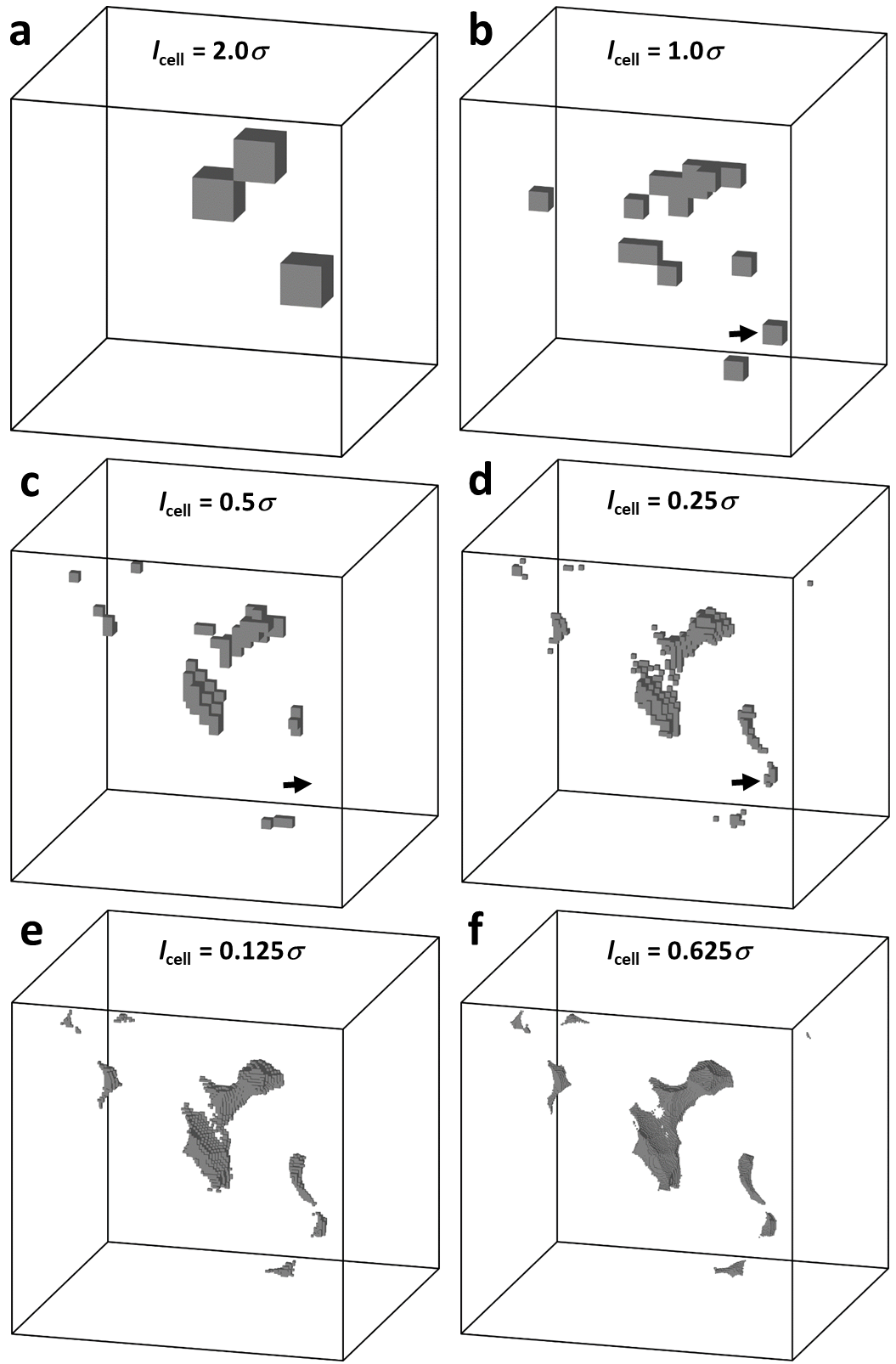}
\caption{\label{fig_snapshots_7} Same as Fig.~\ref{fig_snapshots_17} for another molecular configuration. 
}
\end{figure}

Figures~\ref{fig_snapshots_17} and \ref{fig_snapshots_7} show the result of the application of the W-method on two particular molecular configurations chosen to illustrate different situations.
In each case, the panels (a) to (f) correspond to the six grid meshes considered in this work, ranging from 2$\sigma$ to $2^{-4}\sigma$ respectively. Tables~\ref{table1} and \ref{table2} gather different quantities for the six panels of each figure: the mesh size $l_{\rm cell}$, the number of bubbles $N_B$ in the molecular configuration, and the sizes of the three largest bubbles (when they exist) given in terms of their number of cells $n_i$ or their volume $v_i = n_i \delta$.

\begin{table}
    \centering
    \begin{tabular}{p{3.5em} p{3em} p{3em} p{3em} p{3em} p{3em} p{3em} }
       label                  &  a  &  b   & c     & d     & e     & f      \\
%       [1ex]
%       \vspace{1ex}
       \hline 
       $l_{\rm cell}/\sigma$  & 2.0 & 1.0  & 0.5   & 0.25  & 0.125 & 0.0625 \\
       $N_B$                  & 1   & 1    & 2     & 7     & 11    & 17     \\
       $n_1$                  & 1   & 12   & 70    & 567   & 4552  & 36193  \\
       $n_2$                  & -   & -    & 1     & 16    & 88    & 695    \\
       $n_3$                  & -   & -    & -     & 3     & 28    & 189    \\
       $v_1/\sigma^3$         & 8.0 & 12.0 & 8.75  & 8.859 & 8.891 & 8.836  \\
       $v_2/\sigma^3$         & -   & -    & 0.125 & 0.25  & 0.172 & 0.170  \\
       $v_3/\sigma^3$         & -   & -    & -     & 0.047 & 0.055 & 0.046  \\
    \end{tabular}
    \caption{Grid mesh $l_{\rm cell}$, number of bubbles $N_B$, and sizes of the largest (1), second largest (2) and third largest (3) bubble, given in terms of their number of cells $n_1$, $n_2$ and $n_3$ respectively, or in terms of their volume $v_i = n_i l_{\rm cell}^3$, for the six grid meshes labeled (a) to (f), and for the configuration depicted in Fig.~\ref{fig_snapshots_17}.}
    \label{table1}
\end{table}

\begin{table}
    \centering
    \begin{tabular}{p{3.5em} p{3em} p{3em} p{3em} p{3em} p{3em} p{3em} }
       label                  &  a  &  b   & c     & d     & e     & f      \\
       \hline
       $l_{\rm cell}/\sigma$  & 2.0  & 1.0 & 0.5   & 0.25  & 0.125 & 0.0625 \\
       $N_B$                  & 2    & 6   & 7     & 16    & 9     & 16     \\
       $n_1$                  & 2    & 9   & 39    & 496   & 4052  & 32332  \\
       $n_2$                  & 1    & 3   & 24    & 49    & 390   & 3082   \\
       $n_3$                  & -    & 1   & 4     & 31    & 249   & 2004   \\
       $v_1/\sigma^3$         & 16.0 & 9.0 & 4.875 & 7.75  & 7.91  & 7.89   \\
       $v_2/\sigma^3$         & 8.0  & 3.0 & 3.0   & 0.766 & 0.76  & 0.752  \\
       $v_3/\sigma^3$         & -    & 1.0 & 0.5   & 0.484 & 0.486 & 0.489  \\
    \end{tabular}
    \caption{Same as Table~\ref{table1} for the configuration depicted in Fig.~\ref{fig_snapshots_7}.}
    \label{table2}
\end{table}

For the largest mesh of the first example [Fig.~\ref{fig_snapshots_17}(a)], one observes only one bubble of size one cell, corresponding to a volume $v_1 = 8\sigma^3$. Reducing $l_{\rm cell}$ by a factor two [Fig.~\ref{fig_snapshots_17}(b)] still gives one cluster, the size being now 12 cells and the volume $v_1 = 12\sigma^3$. The shape of the bubble looks irregular and wormlike. For $l_{\rm cell} = 0.5\sigma$ [Fig.~\ref{fig_snapshots_17}(c)], two clusters are observed. The largest one, made of 70 cells, is positioned at the same place as the bubble previously detected in panels (a) and (b), and obviously corresponds to the same bubble seen with a better definition. On the other hand, the second cluster at the bottom of the simulation box contains only one cell and corresponds to a smaller bubble previously undetected in panels (a) and (b). Reducing further the grid mesh allows to detect an increasing number of bubbles in the molecular configuration: for instance, $l_{\rm cell} = 0.0625\sigma$ allows to detect 17 bubbles or cavities [Fig.~\ref{fig_snapshots_17}(f)]. Note however that the newly detected bubbles are quite small, with dimensions of the order of the mesh size that allow to detect them. It is also interesting to notice that, generally, a bubble detected for a given mesh will be detected for smaller meshes. This is in particular the case for the initially detected bubble close to the center of the simulation box [Fig.~\ref{fig_snapshots_17}(a)] which appears in all subsequent panels with a better definition. Obviously, the number of cells in this cluster increases, but the size of the bubble measured by its absolute volume $v_1$ converges (see Table~\ref{table1}). This is also true for the smaller bubbles detected in Fig.~\ref{fig_snapshots_17} (c) to (f). This allows to conclude that the newly detected bubbles in that molecular configuration are 50 times smaller than the largest one. It can also be seen that the shape of the largest bubble converges when the grid mesh decreases. This is also true for the smaller bubbles. 

The second example given in Fig.~\ref{fig_snapshots_7} confirms the previous observations, with however some differences. 
For instance, two clusters are detected for the largest mesh $2\sigma$ [Fig.~\ref{fig_snapshots_7}(a)]. The largest cluster made of two cells corresponds to the largest bubble detected in panel (f), with volume $v_1=7.9\sigma^3$, while the second cluster made of one cell in (a) corresponds to the second largest bubble in (f) with volume $v_2=0.75\sigma^3$. As can be seen, the large grid mesh severely misestimates the volume ratio between the two bubbles: ratio 2 in panel (a), while the ratio is more likely to be 10.5 with a better precision in panel (f) [see Table~\ref{table2}, columns (a) to (f)]. Choosing the standard value $l_{\rm cell} = 0.5\sigma$ instead of $2\sigma$ does not cure the problem since the ratio in that case equals 1.6. Another interesting feature in this configuration is that the number $N_B$ of clusters is not necessarily always increasing when the grid mesh decreases, as can be seen in Table~\ref{table2}, columns (d) and (e). This is due to the merger of clusters close to each other. In particular, one can notice that the two large clusters in panel (c) actually belong to the same cavity when detected with a better accuracy in panel (f). The last point to be noticed is that the single cell cavity pointed out by the arrow in panel (b) disappears in panel (c), and reappears in subsequent panels. This reveals that cavities of a given size may be undetected when analyzed on a grid with a mesh of comparable size, depending on the position and shape of the cavity with respect to the grid.      

These two examples show that the number of cavities detected with the W-method strongly depends on the grid mesh. It is however noticed that the smaller the mesh, the better the accuracy. In particular, the two previous examples show that the largest bubbles are accurately described (in volume and shape) when using the smallest mesh. It looks reasonable to think that whatever the minimal bubble size we want to characterize, it is possible to chose a grid mesh small enough to detect them. In other words, reducing the mesh to zero should allow to characterize all bubbles. However, there remain two important questions. 

(i) From a fundamental point of view, it would be desirable that the method somehow converges, in the sense that the characterization of each configuration should reach a definite limit for a grid mesh approaching zero. In particular, the number of detected cavities should converge toward a finite value. 

(ii) From a practical point of view, it would be desirable that an accurate enough characterization can be reached with a ``reasonable'' value of the grid mesh, in particular in terms of computing resources: as a matter of fact, reducing the grid mesh is highly demanding in memory and computing time, in particular for large simulation boxes. Taking this constrain into account, is the generally admitted $l_{\rm cell} = 0.5\sigma$ the best choice?

To answer these questions, the bubble size distribution is carefully analyzed as a function of the grid mesh.

\subsection{Bubble size distribution}
We first focus on the bubble size distribution with the size defined as the number $n$ of cells in the clusters. By construction, $1 \leq n \leq V/\delta$ where V is the volume of the simulation box and $\delta = l_{\rm cell}^3$ is the cell volume. One first calculates the frequency histogram, $H_\delta(n)$, cumulated over the configurations generated by the simulation and taken with an equal weight. This is equivalent to establishing the histogram based on the ensemble of all cavities detected during the simulation. The histogram is then normalized to one to get $h_\delta(n) = H_\delta(n)/\Sigma_i H_\delta(i)$ which can be interpreted has a probability distribution. 

Since distant bubbles in the simulation box are uncorrelated, this ensemble average is equivalent to considering an infinite system, except for the fact that the histogram is truncated at the upper bound $V/\delta$ that explicitly depends on the system size. This limit can be pushed as far as allowed by computer capabilities, using large enough simulation boxes. On the other hand, the first bin $h_\delta(0)$ cannot be defined and remains inaccessible to numerical calculations since the smallest bubble detected by the W-method contains at least one cell.  

\begin{figure}[t]
\includegraphics[width=0.9\columnwidth]{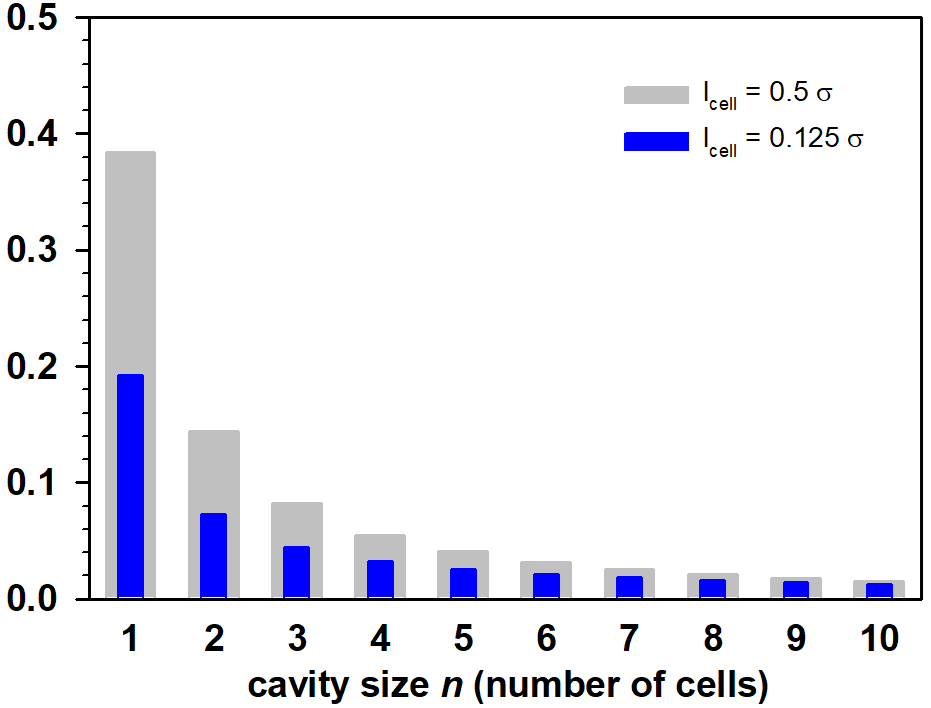}
\caption{\label{fig_histograms_n} Normalized frequency histograms $h_\delta(n)$ of the bubble size measured as the number $n$ of cells in the cluster. Large gray bars: $l_{\rm cell} = 0.5\sigma$; thin blue bars: $l_{\rm cell} = 0.125 \sigma$.}
\end{figure}

The bubble size histogram $h_\delta$ obviously depends on the grid mesh, as can be seen on Fig.~\ref{fig_histograms_n} that gives the result for the two grid meshes $l_{\rm cell} = 0.5\sigma$ and $0.125\sigma$.
In both cases, the probability decreases monotonously when the bubble size increases. The highest probability is for the first bin (single cell cavities), reaching 38\% for $l_{\rm cell} = 0.5\sigma$ and 19\% for $0.125\sigma$, meaning that a large portion of detected cavities have dimensions close to the grid mesh.

\begin{figure}[b]
\includegraphics[width=0.9\columnwidth]{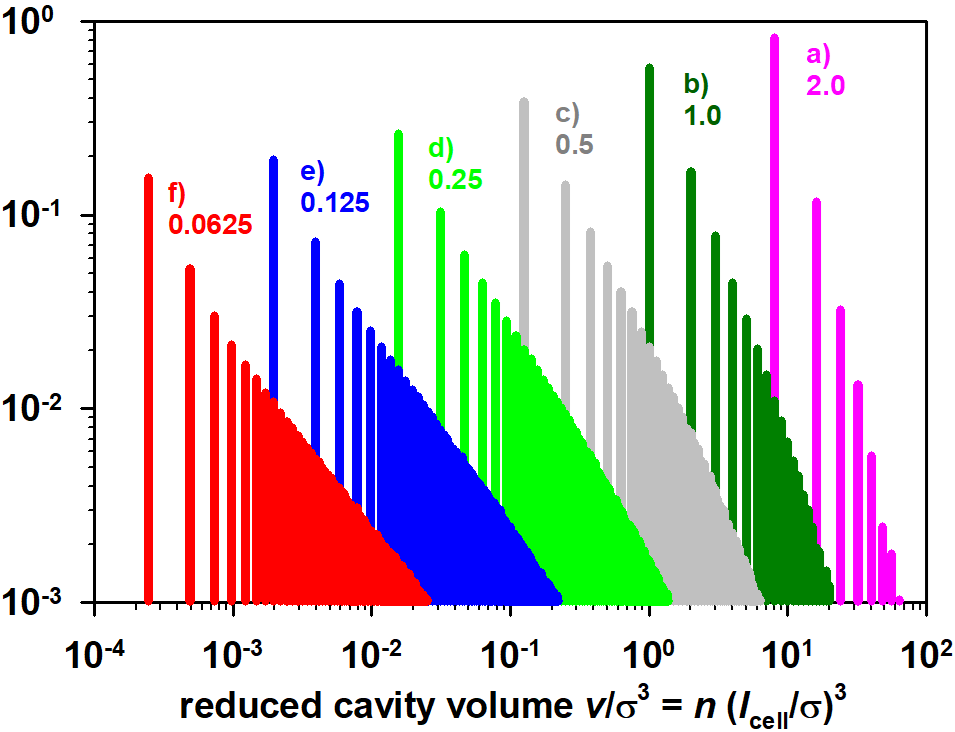}
\caption{\label{fig_histograms_v} Normalized frequency histograms $h_\delta(v)$ given as a function of the reduced cavity volume $v/\sigma^3$ for the six grid meshes $l_{\rm cell}$ given in the figure.}
\end{figure}

Comparison between these normalized histograms requires to represent them as a function of the absolute bubble volume $v = n \delta$. The result is shown in Fig.~\ref{fig_histograms_v} for the six grid meshes considered in this study.  
Because of the large range covered by the bubble volume and probability, the histograms are given in a log-log scale.
Whatever the grid mesh, all histograms exhibit a monotonously increasing probability toward smaller cavities. The smallest grid mesh [Fig.~\ref{fig_histograms_v}(f)] suggests that the probability should follow a power law $h(v)\sim v^{-\alpha}$ for small cavity sizes. This is a puzzling result to be discussed below.
Visual inspection of the evolution of the histograms with grid mesh does not exhibit anything special with the particular value $l_{\rm cell} = 0.5\sigma$ commonly adopted in literature.\cite{RN1184, RN2847} In particular, it is evident that a lot of information below that size is lost.

\begin{figure}[]
\includegraphics[width=0.9\columnwidth]{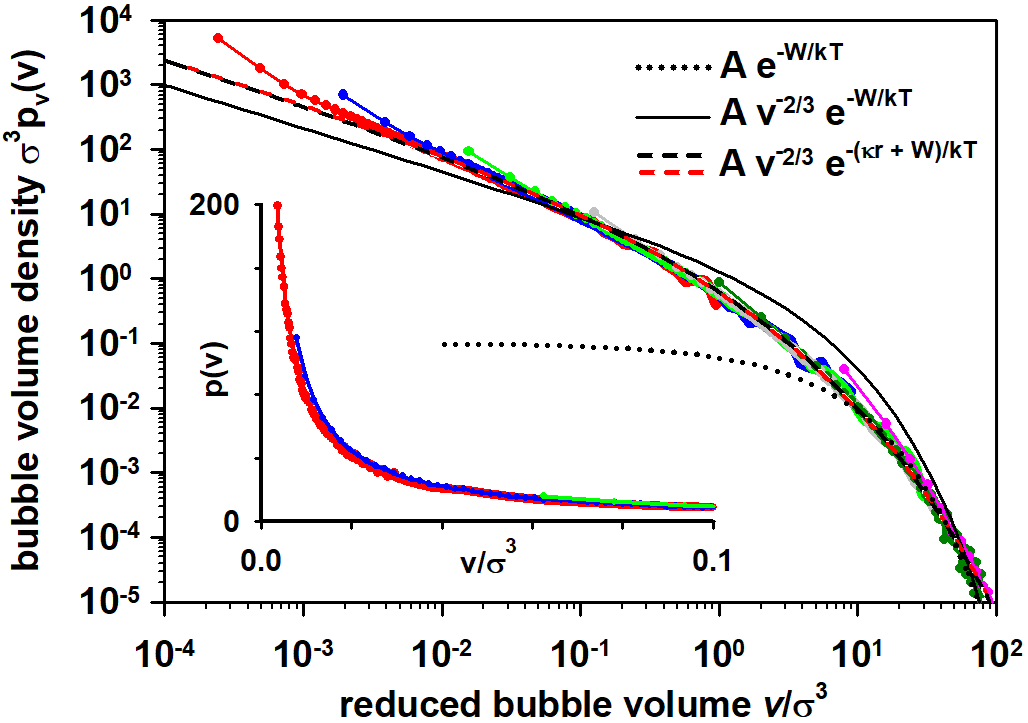}
\caption{\label{fig_p_vs_v_grid} Symbols: reduced bubble size densities $\sigma^3 p_v(v)$ in log-log scale deduced from the histograms $h_\delta(v)$ shown in Fig.~\ref{fig_histograms_v}, and given as a function of the reduced bubble volume $v/\sigma^3$. The inset shows, in linear scale, the same data where the first three bins have been removed. The color code identifying each grid mesh is the same as in Fig.~\ref{fig_histograms_v}. The thin lines connecting the symbols are guides to the eye. The black dotted line is a fit with Eq~\ref{eq_boltz_ps} with constant $Q$; the black solid line is a fit with Eq~\ref{eq_fit_p_1} and the dashed lines are fits with Eq~\ref{eq_fit_p_2} including all data (black) or only the data for $l_{\rm cell} = 0.5 \sigma$ (red).}
\end{figure}

We are now in a position to calculate the density $p_v(v) = dN_v/dv$ using the finite difference approximation $p_v(v) \simeq \Delta N_v/\Delta v$ where $\Delta N_v$ is the number of bubbles with volume between $v$ and $v+\Delta v$. Taking the natural choice $\Delta v = \delta$, $\Delta N_v$ is given by $H_\delta(n)$ with $n=v/\delta$ and $p_v(n\delta) = H_\delta(n)/\delta$ for $1 \leq n\leq V/\delta$. The approximation should be better for bubble sizes much larger than the grid mesh, while it is obviously not appropriate for smaller bubbles.
The result, given in Fig.~\ref{fig_p_vs_v_grid}, shows that all histograms fall on the same curve independent of $\delta$.
It is however interesting to notice that each choice of a grid mesh gives an optimal volume interval for the calculation of $p_v(v)$, between few $\delta$ and hundreds of $\delta$. Below few $\delta$, the grid mesh is too rough to capture the exact volume of the bubble. This can be seen in Fig.~\ref{fig_p_vs_v_grid} where the very first bins of each histogram depart from the main curve. Above hundreds of $\delta$, the bins are too thin, which induces increasing fluctuations that need to be smoothed out. The inset of Fig.~\ref{fig_p_vs_v_grid} shows the data for the three smallest grids in linear scale, beyond the third bin where data are reliable.  

\section{Discussion}
\subsection{The vanishing bubble limit}
How does the numerical results for $p_v(v)$ compare to Eq~\ref{eq_boltz_ps}? At first glance, the noticeable point is that Eq~\ref{eq_boltz_ps} leads to a finite value for $p_v(0)$, while we observe an apparent divergence in Fig.~\ref{fig_p_vs_v_grid}. This is highlighted in the inset where the data obtained for the three smallest grid are reported in linear scale. In order to establish the expected behavior of the bubble size density $p_v(v)$ for vanishing $v$, an analytical calculation has been attempted. We first notice that for a vanishing grid mesh, the voids detected with the W-method converge to the regions out of the Stillinger spheres (corresponding to the white regions in Fig.~\ref{fig_grid}). Therefore, the limit of the distribution $p_v(n\delta) = H_\delta(n)/\delta$ for vanishing $\delta$ can be obtained by calculating the density $p_v(v) = dN_v/dv$ of the distribution of these empty regions (inversion of limits). As we are interested in the behavior of vanishing bubbles, we focus on the tiny empty voids that may appear between the Stillinger spheres. Figure~\ref{fig_2D_3disks_1} gives an example in two-dimensions where a small triangular void is delimited by the Stillinger spheres of three molecules. The Appendix gives the detailed calculation of the volume expansion around zero of $p_v(v)$, in one, two and three dimensions, and proposes a generalization in dimension $d$ (Eq~\ref{eq_p_dD}). In three dimensions, one has:
\begin{equation}
p_v(v)
{\underset{\; v \to 0 \;\,}{=}}
\frac{A_v}{v^{2/3}} + \frac{B_v}{v^{1/3}} + C_v + o(1) \label{eq_DL_p}
\end{equation}
The details of the fluid-fluid interactions are absorbed into the constants $A_v$, $B_v$ and $C_v$. 

In order to make Eqs~\ref{eq_w_cnt}, \ref{eq_boltz_ps} and \ref{eq_DL_p} mutually compatible, $Q$ cannot be a constant and has to vary around $v=0$ according to $Q = A_v v^{-2/3}$. Discarding in Eq~\ref{eq_w_cnt} the negligible volume contribution to the free energy compared to the surface term (we are considering small bubbles), the expression for $p_v(v)$ reads:
\begin{equation}
p_v(v) = \frac{A_v}{v^{2/3}} e^{-(36\pi v^2)^{1/3} \gamma/kT}  \label{eq_fit_p_1}
\end{equation}
where we have made the hypothesis that the bubbles are spherical.

A best fit with the distributions calculated from simulations gives $A_v = 2.15 \sigma^{-1}$, and $\gamma = 0.108 \epsilon/\sigma^2$ (solid black line in Fig.~\ref{fig_p_vs_v_grid}). As can be seen, the agreement is considerably improved compared to a fit with Eq~\ref{eq_boltz_ps} where $Q$ is supposed to be constant (dotted line). This is particularly true on the leftmost side of the figure where the slope of $v^{-2/3}$ clearly accommodates the data. 

On the other hand, for $v\geq 0.5\sigma^3$, the agreement is quite poor. The surface contribution in the exponent varies too rapidly to fit the data. Introducing an asphericity of the bubble would result in an effective surface tension without changing the fit.\cite{RN3194} Restoring the volume contribution in the exponent would not help since the variation would be even faster. Taking into account the term $B_v v^{-1/3}$ from Eq~\ref{eq_DL_p} into the prefactor of Eq~\ref{eq_fit_p_1} does not improve neither. The most efficient way to improve the fit is to introduce a slowly varying contribution $v^{1/3}$ in the exponent:\cite{RN3197, RN3147, RN3183} 
\begin{equation}
p_v(v) = \frac{A_v}{v^{2/3}} e^{-[(36\pi v^2)^{1/3} \gamma + \kappa_v v^{1/3}]/kT}  \label{eq_fit_p_2}
\end{equation}
A best fit of simulation data with Eq~\ref{eq_fit_p_2} gives $A_v = 5.8 \sigma^{-1}$, $\gamma = 5\times 10^{-3} \epsilon/\sigma^2$ and $\kappa_v = 2.2 \epsilon/\sigma$ (black dashed line in Fig.~\ref{fig_p_vs_v_grid}). Note that the obtained value for the surface tension is quite small and the corresponding term plays actually a minor role. Omitting this surface tension term gives essentially the same result with indistinguishable fits in Fig~\ref{fig_p_vs_v_grid}: $A_v = 6.0 \sigma^{-1} $ and $\kappa_v = 2.3 \epsilon/\sigma$. As a matter of fact, the new term $\kappa_v v^{1/3}$ largely dominates for small bubbles with $v \leq 50 \sigma ^3$. A physical interpretation of this term will be given below when larger bubbles will be considered.
The last point to be mentioned is that performing the same fit with only the data for $l_{\rm cell} = 0.5 \sigma$ gives indistinguishable results (red dashed line in Fig~\ref{fig_p_vs_v_grid}). The reason is that the information gained with smaller meshes is entirely contained in the $v^{-2/3}$ dependence for vanishing bubbles. This is an important observation regarding the practical use of the W-method: combined with Eq~\ref{eq_fit_p_2}, $l_{\rm cell}=0.5\sigma$ is a good compromise between precision and computational cost.

\subsection{Volume versus radius densities and free energy profile}
\begin{figure}[t]
\includegraphics[width=0.9\columnwidth]{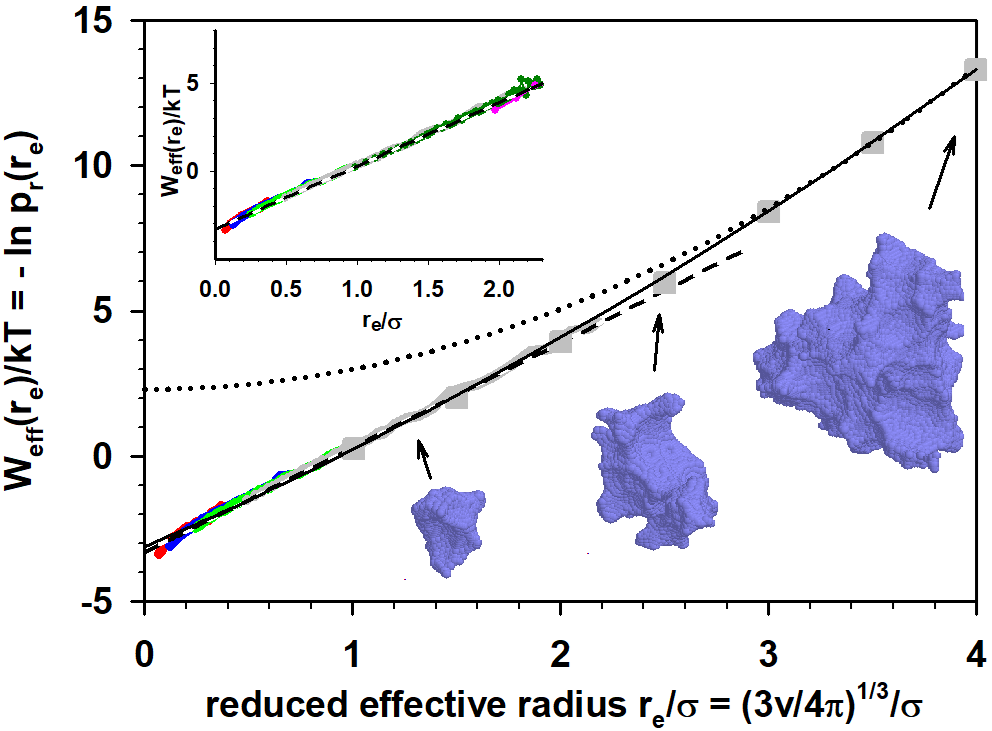}
\caption{\label{fig_p_vs_r_grid} Effective free energy profiles $W_{\rm eff}(r_e)/kT$ given by Eq~\ref{eq_W_eff}, as a function of the bubble effective radius $r_e/\sigma=(3v/4\pi)^{1/3}\sigma^{-1}$. The data (colored symbols) are deduced from Fig.~\ref{fig_p_vs_v_grid} and Eq~\ref{eq_pr_pv}. The color code identifying each grid mesh is the same as in Fig.~\ref{fig_p_vs_v_grid}. The first three bins of each histogram have been omitted. In the inset, all data for $r_e \leq 2\sigma$ are shown, while in the main panel, data for $l_{\rm cell} > 0.5\sigma$ are omitted for clarity. The results from biased umbrella sampling for $r_e \geq \sigma$ are shown as squares in the main panel. The lines are best fits with Eq~\ref{eq_fit_pr} (solid line), or without $W(r_e)$ (dashed line), or with $\kappa_r=0$ (dotted line). Typical bubbles are shown for $r_e/\sigma= 1.3$, 2.5 and 4.0.}
\end{figure}

In most cases, when the capillarity approximation is invoked and the bubble is supposed to be spherical, its size is characterized by its radius instead of its volume. From Eqs~\ref{eq_pr_pv} and \ref{eq_fit_p_2} it is seen that the prefactor in this radial representation is now expected to be a constant, and the new term $\kappa_v v^{1/3}$ is now linear with $r$. In the general case where bubbles are not spherical, one can introduce an effective radius $r_e = (3v/4\pi)^{1/3}$. The corresponding expression for the radius density reads: 
\begin{equation}
p_r(r_e) = A_r e^{-[\kappa_r r_e + W(r_e)]/kT}  \label{eq_fit_pr}
\end{equation}
where $A_r = (36\pi)^{1/3}A_v$, $\kappa_r = (4\pi/3)^{1/3} \kappa_v$, and $W(r_e)$ is the capillarity approximation given by Eq~\ref{eq_w_cnt} where $r$ is taken equal to $r_e$. This equation allows to define an effective free energy 
\begin{equation}
W_{\rm eff}(r_e) = - kT \ln p_r(r_e)  \label{eq_W_eff}
\end{equation}
cleared from the spurious divergence for vanishing bubbles. This can be seen in the inset of Fig.~\ref{fig_p_vs_r_grid} giving $W_{\rm eff}/kT$ calculated from the simulation data.
As previously, the first bins of the histograms are omitted since the corresponding densities deduced from finite difference are erroneous, see Fig.~\ref{fig_p_vs_v_grid}. As can be seen, all histograms fall on the same curve. But now, the noticeable point is that the plotted quantity clearly reaches a finite value at vanishing bubble size. It is possible to define and calculate the limit $W_{\rm eff}(r_e \rightarrow 0)$ (=-$3.3 kT$ in our case) which corresponds to $-kT \ln A_r$. This allows to unambiguously define a free energy difference $W_{\rm eff}(r_e)-W_{\rm eff}(0)$ that can be interpreted as the free energy of formation of the bubble (that correctly reaches zero for a vanishing bubble). It is emphasized that this is not possible if one uses $p_v$ instead of $p_r$ in Eq~\ref{eq_W_eff}.

\subsection{Free energy of formation of a bubble beyond the capillarity approximation}
How does the previous definition of the free energy of formation of a bubble $W_{\rm eff}(r_e)-W_{\rm eff}(0)$ compare with the capillarity approximation? We have already seen that an extra term $\kappa_r r_e$ has to be introduced to fit the data for small bubbles. What happens for larger bubbles?

We have used the umbrella sampling method to calculate the probability distribution up to $r_e = 4 \sigma$.\cite{RN2885} The results are given in Fig.~\ref{fig_p_vs_r_grid} (squares), as well as the previous data for $l_{\rm cell} \leq 0.5\sigma$. The previous fit with Eq~\ref{eq_fit_p_2} and $\gamma=0$ (dashed line in Fig~\ref{fig_p_vs_v_grid}) is also reported in Fig~\ref{fig_p_vs_r_grid} (dashed line, corresponding to Eq~\ref{eq_fit_pr} with $W(r_e)$ set to zero).

As can be seen, the simulation data for large bubbles ($r_e \geq 2 \sigma$) now depart from the dashed straight line corresponding to the previous fit without the surface and volume contributions. Re-introducing the surface tension allows to fit nicely the data over the whole available range (solid line in Fig~\ref{fig_p_vs_r_grid}, with $\gamma = 0.02\epsilon/\sigma^2$; the volume term was unnecessary, but would undoubtedly contribute for larger bubbles). For comparison, a best fit with Eq~\ref{eq_fit_pr} with $\kappa_r=0$ (dotted line) has also been attempted. This would correspond to considering only the usual capillarity approximation. As can be seen, the agreement is quite poor. 

What is the significance of the introduction of the linear term $\kappa_r r_e$? The first point to be mentioned is that the cavity distribution may be method-dependent, in particular for small bubbles. This could influence the magnitude of the linear term: further work is required to evaluate this point. However, in the case of the W-method, a linear term is required to fit the data down to the smallest bubbles. It might be tempting to interpret this term as a consequence of the dependence of the surface tension on the bubble curvature $-2/r$, according to $\gamma(r) = \gamma_\infty/(1-2\tau /r)$, where $\tau \simeq -0.1\sigma$ is Tolman's length.\cite{RN629, RN1293, RN2758, RN3154, RN3182, RN2996} In this context, if one identifies the linear term $\kappa_r r$ with $8\pi \tau \gamma_\infty r$ appearing in the expansion for large bubbles of the surface energy $4\pi r^2 \gamma(r)$, one gets $\tau = \kappa_r/(8\pi \gamma_\infty) = 1.2\sigma$. This is at odds both in sign and magnitude with the expected value. As a matter of fact, this is not contradictory since Tolman's expansion is for \emph{large} bubbles while $\kappa_r r$ is introduced for very small ones. From a physical point of view, $\kappa_r r$ is more likely interpreted as an effective term to take into account the fact that small bubbles are not of spherical shape (see snapshots in Fig~\ref{fig_p_vs_r_grid}). $\kappa_r$ would correspond to an additional linear energetic contribution associated with surface corrugation.

What is the consequence of neglecting the linear term $\kappa_r$ in the context of phase transitions, in particular for the calculation of nucleation barriers? By definition, the barrier is the free energy of formation of the critical nucleus, i.e. $W(r_c) - W(0)$ where $r_c$ is the critical radius. If the linear contribution is omitted in Eq~\ref{eq_fit_pr} ($\kappa_r = 0$), i.e. $W_{\rm eff}$ is replaced by the capillarity approximation Eq~\ref{eq_w_cnt}, a best fit can be performed on the largest bubbles where this approximation is expected to be better ($r_e \geq 3 \sigma$ in our case). The result is given in Fig~\ref{fig_p_vs_r_grid} (dotted line). As can be seen, this results in a significant difference for $W(0)$, 5.5$kT$ in our case, that reports directly on the nucleation barrier. For comparison, the CNT predicts a reduced barrier around 20$kT$ for $kT/\epsilon = 1$ and $\Delta P \sigma ^3 / \epsilon = -0.03$ corresponding to our thermodynamic conditions: the error then reaches 25\%. For larger barriers, more likely to appear in realistic experimental situations, the relative error would be smaller, but not necessarily negligible.

\section{Conclusion} 
Spontaneous tiny bubbles are ubiquitous in liquids, in the stable as well as in the metastable state. They are particularly important in the context of the liquid-to-vapor phase transition as they play a fundamental role in the process of nucleation. In particular, the probability of finding bubbles of the critical size appears in the classical nucleation theory. More generally, the bubble size distribution is generally translated into a free energy profile through Eq~\ref{eq_boltz_ps}, with an associated free energy of formation that can be written in terms of surface and volume contributions in the capillarity approximation (Eq~\ref{eq_w_cnt}) valid for large bubbles. In this work we have focused on small bubbles or cavities, typically of the order of the molecular size. We have shown that:

(i) The W-method \cite{RN1184} is efficient to detect tiny bubbles if the mesh $l_{\rm cell}$ is chosen small enough. In practice, it is enough to take $l_{\rm cell} = 0.5\sigma$: large bubbles are well characterized, while the size distribution of small bubbles is well accounted for thanks to the following points (ii) and (iv). 

(ii) The bubble volume distribution $p_v$ follows a $v^{-2/3}$ law in the vanishing limit, while the distribution $p_r$ calculated for the equivalent radius $r_e=(3v/4\pi)^{1/3}$ converges toward a constant in virtue of Eq~\ref{eq_pr_pv}.

(iii) One can define an effective free energy according to Eq~\ref{eq_W_eff}. Thanks to point (ii), $W_{\rm eff}(0)$ exists, and it is possible to define the free energy of formation of the bubble as $W_{\rm eff}(r_e)-W_{\rm eff}(0)$, which correctly reaches zero for a vanishing bubble. Despite the fact that, in most cases, Eq~\ref{eq_W_eff} seems to be correctly used in literature, it has never been clearly stated, and implications of Eq~\ref{eq_pr_pv} have never been discussed. It is however emphasized that using $p_v$ instead of $p_r$ in Eq~\ref{eq_W_eff} does not allow to define properly a free energy of formation of the bubble. 

(iv) Comparison of $W_{\rm eff}$ with the capillarity approximation Eq~\ref{eq_w_cnt} shows that it is necessary to introduce a linear term $\kappa_r r_e$, that can be interpreted as an effective correction to the fact that bubbles significantly depart from the spherical shape.

As a conclusion, gathering these recommendations in the context of nucleation should allow to improve the calculation of free energy barriers, in particular when the critical bubble is small and the corresponding barrier is low. 

\begin{acknowledgments}
The author acknowledges fruitful discussions with E. Rolley and P. Porion, and is grateful to the CNRS interdisciplinary “Défi
Needs” through its “MiPor” program (Project DARIUS) and the financial support of Agence Nationale de la Recherche through the project ANR-23-CE30-0028-03 (NanoCav).
\end{acknowledgments}

\section*{Author Declarations}
\subsection*{Conflict of interest}
The author has no conflicts to disclose.
\subsection*{Author contributions}
Jo\"el Puibasset: Conceptualization ; Data curation ;
Writing – original draft ; Writing – review \& editing.

\section*{Data Availability Statement}
The data that support the findings of this study are available from the corresponding author upon reasonable request.

\section{Appendix}
This section is devoted to the calculation of the bubble size distribution in the vanishing limit. It is shown that a general relation can be obtained with minimal hypotheses. For clarity, the argument is first presented in details in dimensions 1 and 2. In dimension 3 we give only the important steps, leading to a generalization to dimension $d$. 

\subsection{Definitions}
The first step of the W-method consists in identifying the liquid-like atoms with the Stillinger's criterion,\cite{RN3144} which compares the number of neighboring molecules with a given threshold. The liquid-like atoms are then replaced by a sphere of radius corresponding to the first minimum of the radial distribution function (Stillinger's sphere). These criteria obviously depend on the space dimension $d$: the radius of the Stillinger $d$-dimensional sphere will be denoted as $R_d$, where $R_3 = 1.625\sigma$ in our case of a Lennard-Jones fluid at $kT/\epsilon = 1$. The positions of the Stillinger spheres are denoted as ${\bf r}_i$, and the distance between spheres $i$ and $j$ is denoted as $r_{ij}$.

Let us now focus on the conditions to have a vanishing bubble, corresponding to the smallest voids that can be built between the Stillinger spheres. In one dimension, these voids consists in linear intervals between two adjacent and non-overlapping 1D-spheres. In two dimensions, they consist in curvilinear triangles formed by three disks which overlap in pairs, but not all three together. In three dimensions, four spheres are required, which overlap three by three, but not all four together. Let us denote $V_d({\bf r}_{i_0},...,{\bf r}_{i_d})$ the volume of the void between the $(d+1)$ $d$-dimensional spheres $i_0, ...,i_d$ verifying the above mentioned conditions. The probability to have a void of volume equal to a given infinitesimal $d$-volume $v_d$ is proportional to the average number of such voids:
\begin{align}
\begin{split}
&p_v(v_d) \propto \\
&\int e^{-H({\bf r}_1,...,{\bf r}_{N})/kT} \sum_{i_0, ..., i_d} \delta(V_d({\bf r}_{i_0},...,{\bf r}_{i_d})-v_d) d{\bf r}_1...d{\bf r}_{N}
\end{split}
\end{align}
where the sum runs over all ($d$+1)-uplet $(i_0, ...,i_d)$ of spheres and $\delta$ is the Dirac distribution that selects those satisfying the condition to form a small void with the given volume $v_d$. The exponential is the Boltzmann weight associated with the statistical ensemble under consideration, chosen to be canonical in this example for simplicity. However, the argument can be generalized to any statistical ensemble. The integration runs over the coordinates of the $N$ spheres in the system. 

The vanishing voids between Stillinger spheres are rare enough to be considered as spatially independent. Therefore, their number distribution follows a Poisson law, and the average number of voids of a given size in proportional to the probability to have exactly one void of the given size. By symmetry, all the ($d$+1)-uplets of the sum contribute equally. Therefore, the calculation can be done only for the particular choice of spheres 1 to ($d$+1): 
\begin{align}
\begin{split}
&p_v(v_d) \propto \\
&\int e^{-H'({\bf r}_1,...,{\bf r}_{d+1})/kT}\delta(V_d({\bf r}_{1},...,{\bf r}_{d+1})-v_d) d{\bf r}_1...d{\bf r}_{d+1} \label{eq_P_vd}
\end{split}
\end{align}
where the integration now runs only on the ($d$+1) positions of the Stillinger spheres that form the void. This is possible through a redefinition of the Boltzmann weight by integration on the remaining spheres:
\begin{equation}
e^{-H'({\bf r}_1,...,{\bf r}_{d+1})/kT} = \int e^{-H({\bf r}_1,...,{\bf r}_{N})/kT} d{\bf r}_{d+2}...d{\bf r}_{N}
\end{equation}

\subsection{In dimensions one and two}
In one-dimensional space, the volume of the void between the two 1D-spheres is $V_1({\bf r}_{1},{\bf r}_{2}) = V_1(r_{12}) = r_{12} - 2R_1$ if the mutual distance $r_{12} \geq 2R_1$ and is zero otherwise. By translation invariance, $H'({\bf r}_1,{\bf r}_2)= H'(r_{12})$, and can be expanded around $2R_1$: $H'({\bf r}_1,{\bf r}_2) = H'(2R_1) + \alpha V_1 + o(V_1)$. Equation \ref{eq_P_vd} then gives $p_v(v_1) \propto Ve^{H'(2R_1)}(1+\alpha v_1 + o(v_1))$, where the total volume $V$ comes from the global translation invariance. The probability is then finished around $v_1=0$.

\begin{figure}[b]
\includegraphics[width=0.9\columnwidth]{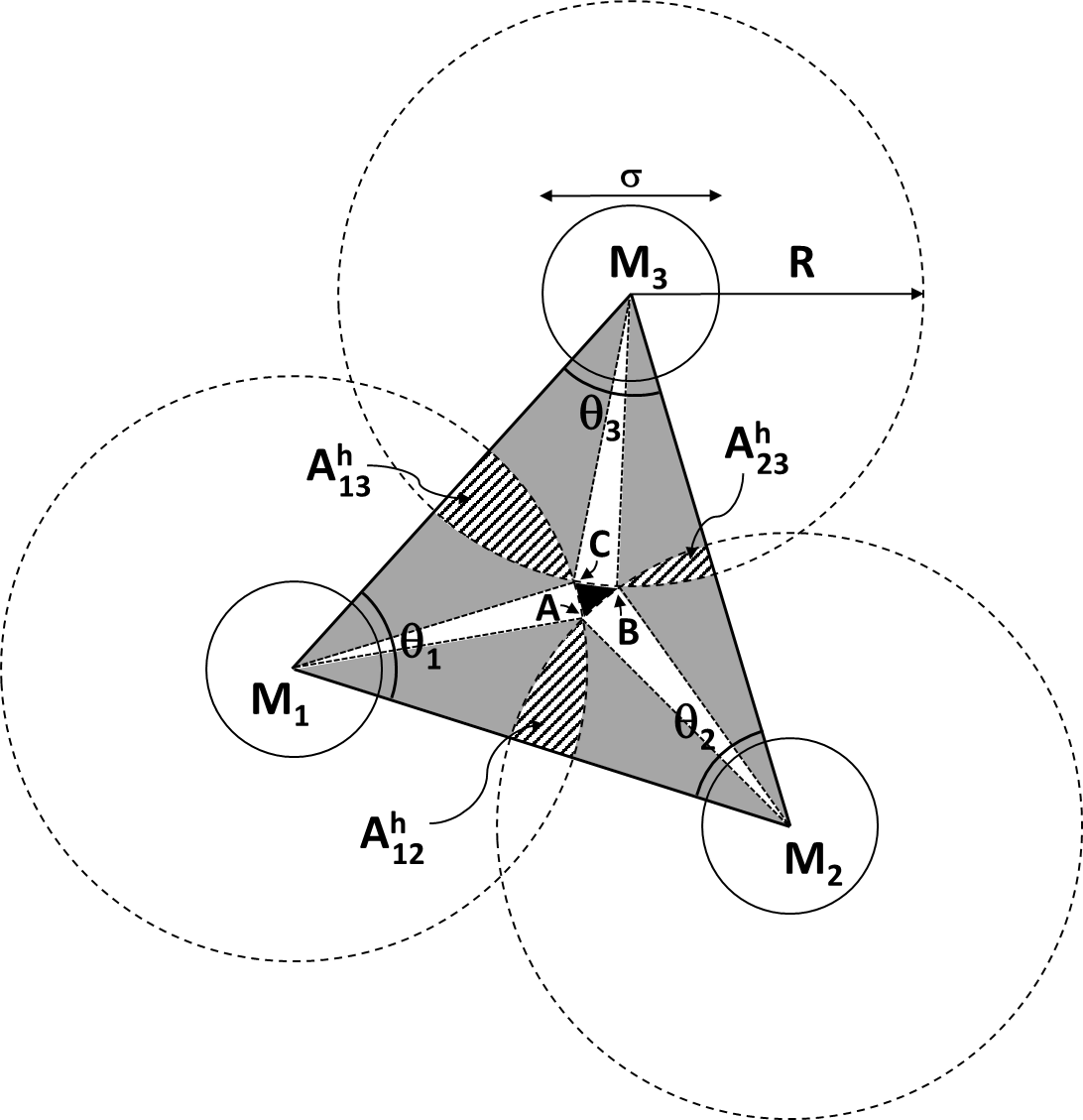}
\caption{\label{fig_2D_3disks_1} Schematic representation in the two-dimensional case of three atoms (solid circles of diameter $\sigma$) at ${\bf r}_{1}$, ${\bf r}_{2}$ and ${\bf r}_{3}$ materialized by the points $M_1$, $M_2$ and $M_3$, and their corresponding Stillinger 2D-spheres represented by the dotted circles. A small void (central curvilinear triangle $ABC$ in black) may exist between these spheres if they overlap by pairs, but do not overlap all three together. The hatched areas resulting from the intersection between disks $i$ and $j$ and the main triangle $M_1 M_2 M_3$ are denoted as $A^h_{ij}$. 
}
\end{figure}

In two-dimensional space, the expression for $V_2({\bf r}_1,{\bf r}_2,{\bf r}_3)$ is more involved, but depends only on the mutual distances: $V_2(r_{12},r_{13},r_{23})$. The first condition, that the 2D-spheres overlap by pairs, is $r_{ij} \leq 2R_2$. Such a situation is depicted in Fig.~\ref{fig_2D_3disks_1}, where the three molecules at ${\bf r}_1,{\bf r}_2,{\bf r}_3$ (materialized by the points $M_1$, $M_2$ and $M_3$) are represented by solid circles of diameter $\sigma$, and their corresponding Stillinger 2D-spheres are represented by the dotted circles.
The second condition, that the three spheres do not overlap together, is equivalent to the existence of the 3-branch star delimited by the points $M_1 A M_2 B M_3 C$, where $A$, $B$ and $C$ are at the intersection between the three Stillinger disks that fall inside the triangle $M_1 M_2 M_3$ (see Fig.~\ref{fig_2D_3disks_1}). The area of this star is given by the area $A_{123}$ of the main triangle $M_1 M_2 M_3$, minus the sum of the areas of the three minor triangles $M_1 A M_2$, $M_2 B M_3$ and $M_1 C M_3$ in gray in Fig.~\ref{fig_2D_3disks_1}. Using Heron's formula, $A_{123} = 0.25 \sqrt{(r_{12}^2+r_{13}^2+r_{23}^2)^2-2(r_{12}^4+r_{13}^4+r_{23}^4)}$ and the sum of the area of the minor triangles is given by $\sum_{ij} 0.25 r_{ij}\sqrt{4R_2^2-r_{ij}^2}$ for $ij = 12$, $23$ and $13$. When these conditions are fulfilled, the void $ABC$ (in black in Fig.~\ref{fig_2D_3disks_1}) exists and its area is calculated as follows. Starting from the area of the full triangle $M_1 M_2 M_3$, one first removes the area of the three circular sectors defined by the intersection of the disks 1, 2 and 3 with the triangle $M_1 M_2 M_3$. This area equals $0.5(\theta_1+\theta_2+\theta_3)R_2^2=0.5\pi R_2^2$ (see Fig.~\ref{fig_2D_3disks_1}). However, each hatched area $A^{h}_{ij}$ in Fig.~\ref{fig_2D_3disks_1} delimited by the intersection between the two disks $i$ and $j$ and the main triangle $M_1 M_2 M_3$ is removed twice, and has to be compensated. Using $A^{h}_{ij} = R_2^2 \arccos(0.5r_{ij}/R_2) - 0.25 r_{ij}\sqrt{4R_2^2-r_{ij}^2}$, the final result reads:
\begin{equation}
V_2({\bf r}_{1},{\bf r}_{2},{\bf r}_{3}) = A_{123} - 0.5\pi R_2^2 + A^{h}_{12} + A^{h}_{23} + A^{h}_{13}. \label{Eq_V2}
\end{equation}
This formula can be used to calculate numerically the exact volume of voids during the simulation of a 2D-fluid without resorting to a grid method.

\begin{figure}[b]
\includegraphics[width=0.9\columnwidth]{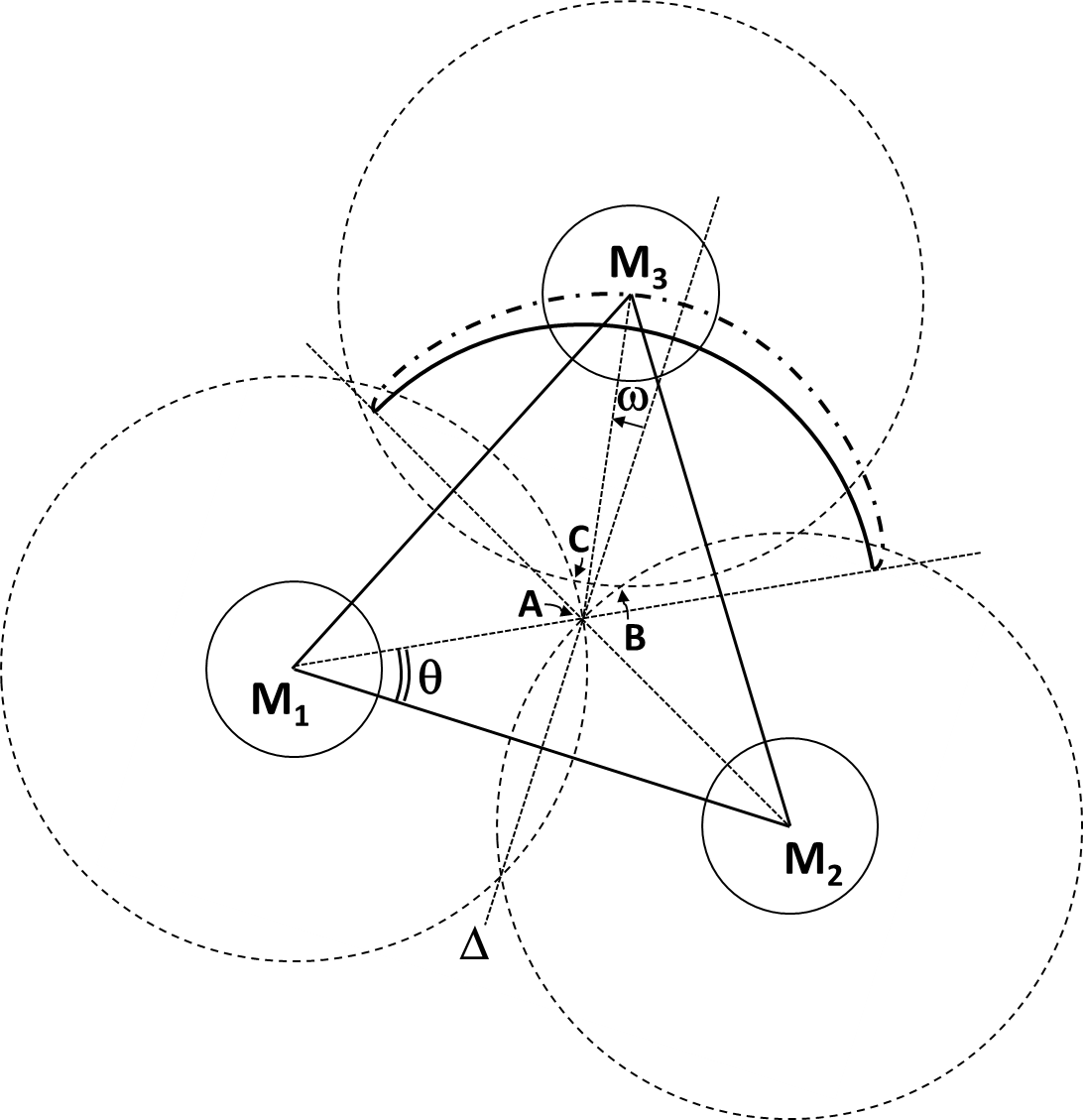}
\caption{\label{fig_2D_3disks_2} Same atomic configuration as in Fig.~\ref{fig_2D_3disks_1}. $\Delta$ is the perpendicular bisector of points $M_1$ and $M_2$. $l_3$ and $\omega$ are the polar coordinates of $M_3$ with respect to point $A$ and axis $\Delta$. The solid circular arc is the locus of $M_3$ giving a vanishing void $V_2$, while the dashed-dotted curve is the locus of $M_3$ giving the finite void volume $V_2$ corresponding to the depicted atomic configuration.   
}
\end{figure}

\begin{figure}[b]
\includegraphics[width=0.9\columnwidth]{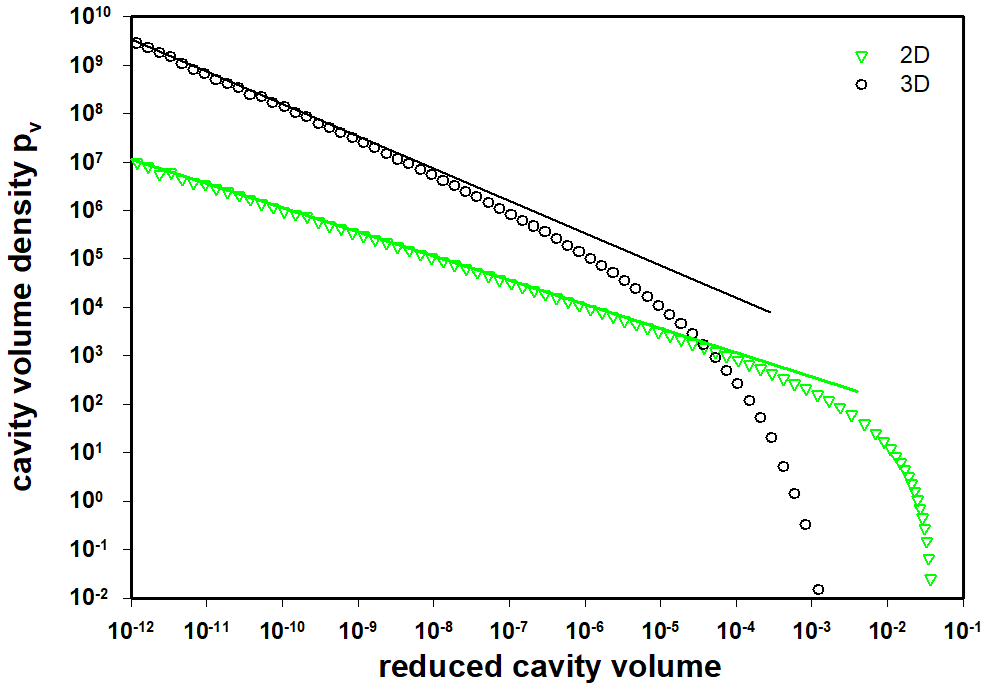}
\caption{\label{fig_p_vs_v_2D_3D} Symbols: cavity size density $p_v$ in log-log scale for random disks in a two-dimensional box (green triangles) or spheres in a three-dimensional box (black circles). The green and black lines correspond to the main term in Eqs~\ref{eq_p_2D} and \ref{eq_p_3D} for the 2D and 3D cases respectively.
}
\end{figure}

The analytical calculation of $p_v(v_2)$ using Eqs~\ref{eq_P_vd} and \ref{Eq_V2} can be done by noticing that $V_2$ is invariant by translation and rotation of the tree atoms as a whole. One then introduces the relative positions ${\bf r'}_{i} = {\bf r}_{i} - {\bf r}_{1}$ for $i$ = 2 and 3. The integration over $d{\bf r'}_{2}$ is donne in polar coordinates with trivial integration on the angular coordinate: $\int d{\bf r'}_{2}$ is replaced by $2\pi\int r_{12}dr_{12}$. The remaining integration on $d{\bf r'}_{3}$ is done as follows. When $r_{12}$ is chosen, the point A in Fig.~\ref{fig_2D_3disks_1} is determined. We introduce polar coordinates centered on A, the angles being measured with reference to the perpendicular bisector of points $M_1$ and $M_2$, denoted as $\Delta$ in Fig.~\ref{fig_2D_3disks_2}.
The polar coordinates of the third atom $M_3$ are denoted as $l_3 = |A M_3|$ and $\omega$ (see Fig.~\ref{fig_2D_3disks_2}). The void $V_2$ exists only if $l_3 \geq R_2$. Let us denote $\xi = l_3 - R_2 \geq 0$. One has: $V_2({\bf r}_{1},{\bf r}_{2},{\bf r}_{3}) = V_2(r_{12},\xi,\omega)$. For a vanishing void, $\xi \rightarrow 0$, and the loci of the points $M_3$ compatible with this condition fall on the circular arc of radius $R_2$, centered on $A$, and limited by the two straight lines $M_1 A$ and $M_2 A$ extended on the side of $M_3$, corresponding to the constraint that the Stillinger disk of the third atom has to cut the two other ones (see the solid circular arc in Fig.~\ref{fig_2D_3disks_2}).
For a finite $V_2 = v_2$, the loci of $M_3$ points can be determined numerically by calculating $\xi^*(v_2,r_{12},\omega)$ such that $V_2(r_{12},\xi^*,\omega)=v_2$ (see for example the dash-dotted line in Fig.~\ref{fig_2D_3disks_2} that gives the loci of points $M_3$ giving the void volume corresponding to the depicted atomic configuration). Gathering everything:
\begin{align}
\begin{split}
& p_v(v_2) \propto \\ &\int e^{-H'(r_{12},\xi,\omega)/kT} \delta(V_2(r_{12},\xi,\omega)-v_2) (\xi+R_2) r_{12} dr_{12} d\omega d\xi
\end{split}
\end{align}
where the integral runs from 0 to 2$R_2$ for $r_{12}$, from 0 to $R_2$ for $\xi$ and from $\theta-\pi/2$ to $\pi/2-\theta$ for $\omega$. The calculation can be done numerically for any void $v_2$, but an analytical expansion can be found in the vanishing limit $v_2 \rightarrow 0$. Expanding Eq~\ref{Eq_V2}, it can easily be shown that    
\begin{equation}
V_2(r_{12},\xi,\omega) = 0.5 \xi^2 \lbrace \tan(\theta+\omega)+\tan(\theta-\omega)\rbrace + o(\xi^2) \label{Eq_dV2}
\end{equation}
where $\theta$ is defined in Fig.~\ref{fig_2D_3disks_2}. Integration of the delta function with respect to $\xi$ gives a factor $(dV_2/d\xi)^{-1} = 0.5 \xi^*(v_2,r_{12},\omega)/v_2$ where $\xi^*$ as been previously defined. One gets:
\begin{equation}
p_v(v_2) \propto \int e^{-H'(r_{12},\xi^*,\omega)/kT} \{ 0.5\xi^*(\xi^*+R_2)/v_2 \} r_{12} dr_{12} d\omega
\end{equation}
Noticing that, from Eq~\ref{Eq_dV2}, $\xi^*(v_2,r_{12},\omega) = f(r_{12},\omega)\sqrt{v_2} + o(\sqrt{v_2})$, and expanding $H'(r_{12},\xi^*,\omega) = H'(r_{12},0,\omega) + \alpha(r_{12},\omega) \xi^* + o(\xi^*)$, the probability can be written as:
\begin{equation}
p_v(v_2) \propto \frac{A_2}{\sqrt{v_2}} + B_2 + o(1) \label{eq_p_2D}
\end{equation}
where 
\begin{equation}
A_2 = 0.5 \int e^{-H'(r_{12},0,\omega)/kT} f(r_{12},\omega) R_2 r_{12} dr_{12} d\omega \label{eq_cte_A}
\end{equation}
and $B_2 =$
\begin{equation}
0.5 \int e^{-H'(r_{12},0,\omega)/kT} f(r_{12},\omega)^2 (1+\alpha(r_{12},\omega) R_2) r_{12} dr_{12} d\omega \label{eq_cte_B}
\end{equation}
Figure~\ref{fig_p_vs_v_2D_3D} shows the results of a two-dimensional simulation of disks in a plane.
We have chosen to ignore the interactions (random disks). This is possible since their influence is limited to the constants $A_2$, $B_2$, etc. appearing in the expansion. The main behavior $\sim 1/\sqrt{v_2}$ is thus independent of the interactions. As can be seen in Fig.~\ref{fig_p_vs_v_2D_3D} the agreement is excellent in the vanishing limit (straight line in log-log scale). 

\subsection{In dimensions three and higher}

In three dimensions, the expression of $V_3$ is even more involved, and an exact expression has not be found. However, as previously, invariance by translation and rotation can be used, as well as an expansion of the void volume in the vanishing limit. The calculation in dimension three essentially proceeds in a way similar to the 2D case, with the following differences:

(i) The conditions to have a closed void between the four Stillinger spheres require that they overlap three by three. In the previous 2D case, we encountered the opposite condition that three disks do not overlap, requiring that the area of the main triangle $M_1 M_2 M_3$ in Fig.~\ref{fig_2D_3disks_1} be larger than the sum of the areas of the three minor triangles $M_1 A M_2$, $M_2 B M_3$ and $M_1 C M_3$ in gray in Fig.~\ref{fig_2D_3disks_1}. This can easily be translated into a condition of overlapping of three spheres by considering their 2D intersection with the plane defined by their centers. Similarly, the condition of non-overlapping of the four spheres imposes a condition on their relative distances that will define the integration limits (next points).

(ii) As previously the integration is done using radial and angular coordinates. The previous integration on $r_{12}$ in 2D polar coordinates is replaced by a 3D spherical integration with the measure $r_{12}^2 dr_{12}$.

(iii) The position of the third atom is measured with respect to the first two ones, in cylindrical coordinates ($r_3$,$\phi_3$,$z_3$) around the $z$ axis defined by the two first atoms. $r_3$ is the distance of atom 3 with respect to this axis. The corresponding integration measure is $r_3 dr_3 d\phi_3 dz_3$. By rotation invariance, integration over $\phi_3$ is trivial.

(iv) The position of the fourth atom is measured in spherical coordinates with the origin at the intersection of the first three spheres. As in the 2D case, a void between the four spheres exists only if the radial coordinate $r_4$ is larger than $R_3$, the Stillinger radius in dimension three. We thus introduce $\xi = r_4 - R_3 \geq 0$. The orientation is measured with two angles denoted $\omega_4$ and $\phi_4$, and the integration measure is $ (\xi + R_3)^2 d\xi \sin{\omega_{4}} d\omega_{4} d\phi_{4}$. 

The probability can therefore be written as follows:
\begin{equation}
p_v(v_3) = \int e^{-H'(\xi,\Omega)/kT} \delta(V_3(\xi,\Omega)-v_3) (\xi+R_3)^2 d\xi d\Omega
\end{equation}
where $\Omega$ is a shortcut for $(r_{12},r_3,z_3,\omega_4,\phi_4)$ and $d\Omega = \sin{\omega_{4}} dr_{12} dr_{3} dz_{3} d\omega_{4} d\phi_{4}$. 

Simple geometric considerations give the expansion of $V_3$ in the vanishing limit:
\begin{equation}
V_3(\xi,\Omega) = \xi^3 g(\Omega) \label{Eq_dV3}
\end{equation}
We introduce $\xi^*(v_3,\Omega)$ such that $V_3(\xi^*,\Omega) = v_3$. The delta function introduces the factor $(dV_3/d\xi)^{-1} = (1/3) \xi^*/v_3$, and one finally gets:
\begin{equation}
p_v(v_3) \propto \int e^{-H'(\xi^*,\Omega)/kT} \{ (1/3)\xi^*(\xi^*+R_3)^2/v_3 \} d\Omega
\end{equation}
As previously, expanding $H'(\xi,\Omega)$ around $\xi=0$, one finally gets 
\begin{equation}
p_v(v_3) \propto \frac{A_3}{v_3^{2/3}} + \frac{B_3}{v_3^{1/3}} + C_3 + o(1) \label{eq_p_3D}
\end{equation}
where $A_3$, $B_3$ and $C_3$ follow from equations similar to Eqs~\ref{eq_cte_A} and \ref{eq_cte_B}.

Figure~\ref{fig_p_vs_v_2D_3D} shows the result of a three-dimensional simulation of spheres in space. As previously, the interactions are ignored since their influence is limited to the constants $A_3$, $B_3$, etc. As can be seen, the distribution in the vanishing limit follows the $v_3^{-2/3}$ law (straight line in log-log scale).

The previous argument suggests a generalization to dimension $d$. In particular, we use the symmetries as well as the radial and angular coordinates. Dimensional argument is used to obtain the cavity expansion in the vanishing limit resulting in an equation similar to Eq~\ref{Eq_dV3}. After integration, one gets:
\begin{equation}
p_v(v_d) \propto \frac{A_d}{v_d^{1-1/d}} + \frac{B_d}{v_d^{1-2/d}} + \frac{C_d}{v_d^{1-3/d}} + ... \label{eq_p_dD}
\end{equation}
where the details of the interatomic forces are absorbed into the constants $A_d$, $B_d$, etc. As can be seen, except in one-dimension, the probability diverges for vanishing voids.

In the context of the classical nucleation theory, the bubbles are supposed to be spherical, and the probability distribution of bubble sizes is generally given as a function of the sphere radius $r$. In this case, the bubble size density should verify $p_r(r)dr = p_v(v_d)dv_d$ (= the number of bubbles between $r$ and $r+dr$). Since $v_d \propto r^d$, one has:
\begin{equation}
p_r(r) = d \frac{v_d}{r} p_v(v_d) \propto A_d' + B_d' r + C_d' r^2 + ...
\end{equation}
In this radial representation, the bubble size density is expected to be finite for vanishing bubbles. This is an important point that allows to define the free energy of formation of a bubble (see main text).

%\section*{References}
\section*{}
\nocite{*}
\bibliography{bubble_stat_size_grid.bib}% Produces the bibliography via BibTeX.

%aipnum4-2.bst 2019-01-14 (MD) hand-edited version of apsrev4-1.bst
\begin{thebibliography}{72}%
\makeatletter
\providecommand \@ifxundefined [1]{%
 \@ifx{#1\undefined}
}%
\providecommand \@ifnum [1]{%
 \ifnum #1\expandafter \@firstoftwo
 \else \expandafter \@secondoftwo
 \fi
}%
\providecommand \@ifx [1]{%
 \ifx #1\expandafter \@firstoftwo
 \else \expandafter \@secondoftwo
 \fi
}%
\providecommand \natexlab [1]{#1}%
\providecommand \enquote  [1]{``#1''}%
\providecommand \bibnamefont  [1]{#1}%
\providecommand \bibfnamefont [1]{#1}%
\providecommand \citenamefont [1]{#1}%
\providecommand \href@noop [0]{\@secondoftwo}%
\providecommand \href [0]{\begingroup \@sanitize@url \@href}%
\providecommand \@href[1]{\@@startlink{#1}\@@href}%
\providecommand \@@href[1]{\endgroup#1\@@endlink}%
\providecommand \@sanitize@url [0]{\catcode `\\12\catcode `\$12\catcode `\&12\catcode `\#12\catcode `\^12\catcode `\_12\catcode `\%12\relax}%
\providecommand \@@startlink[1]{}%
\providecommand \@@endlink[0]{}%
\providecommand \url  [0]{\begingroup\@sanitize@url \@url }%
\providecommand \@url [1]{\endgroup\@href {#1}{\urlprefix }}%
\providecommand \urlprefix  [0]{URL }%
\providecommand \Eprint [0]{\href }%
\providecommand \doibase [0]{https://doi.org/}%
\providecommand \selectlanguage [0]{\@gobble}%
\providecommand \bibinfo  [0]{\@secondoftwo}%
\providecommand \bibfield  [0]{\@secondoftwo}%
\providecommand \translation [1]{[#1]}%
\providecommand \BibitemOpen [0]{}%
\providecommand \bibitemStop [0]{}%
\providecommand \bibitemNoStop [0]{.\EOS\space}%
\providecommand \EOS [0]{\spacefactor3000\relax}%
\providecommand \BibitemShut  [1]{\csname bibitem#1\endcsname}%
\let\auto@bib@innerbib\@empty
%</preamble>
\bibitem [{\citenamefont {Kirkwood}\ and\ \citenamefont {Buff}(1951)}]{RN2963}%
  \BibitemOpen
  \bibfield  {author} {\bibinfo {author} {\bibfnamefont {J.~G.}\ \bibnamefont {Kirkwood}}\ and\ \bibinfo {author} {\bibfnamefont {F.~P.}\ \bibnamefont {Buff}},\ }\bibfield  {title} {\enquote {\bibinfo {title} {The statistical mechanical theory of solutions. {I}},}\ }\href {https://doi.org/10.1063/1.1748352} {\bibfield  {journal} {\bibinfo  {journal} {The Journal of Chemical Physics}\ }\textbf {\bibinfo {volume} {19}},\ \bibinfo {pages} {774--777} (\bibinfo {year} {1951})}\BibitemShut {NoStop}%
\bibitem [{\citenamefont {Reiss}, \citenamefont {Frisch},\ and\ \citenamefont {Lebowitz}(1959)}]{RN2859}%
  \BibitemOpen
  \bibfield  {author} {\bibinfo {author} {\bibfnamefont {H.}~\bibnamefont {Reiss}}, \bibinfo {author} {\bibfnamefont {H.~L.}\ \bibnamefont {Frisch}},\ and\ \bibinfo {author} {\bibfnamefont {J.~L.}\ \bibnamefont {Lebowitz}},\ }\bibfield  {title} {\enquote {\bibinfo {title} {Statistical mechanics of rigid spheres},}\ }\href {https://doi.org/10.1063/1.1730361} {\bibfield  {journal} {\bibinfo  {journal} {The Journal of Chemical Physics}\ }\textbf {\bibinfo {volume} {31}},\ \bibinfo {pages} {369--380} (\bibinfo {year} {1959})}\BibitemShut {NoStop}%
\bibitem [{\citenamefont {Helfand}\ \emph {et~al.}(1960)\citenamefont {Helfand}, \citenamefont {Reiss}, \citenamefont {Frisch},\ and\ \citenamefont {Lebowitz}}]{RN2875}%
  \BibitemOpen
  \bibfield  {author} {\bibinfo {author} {\bibfnamefont {E.}~\bibnamefont {Helfand}}, \bibinfo {author} {\bibfnamefont {H.}~\bibnamefont {Reiss}}, \bibinfo {author} {\bibfnamefont {H.~L.}\ \bibnamefont {Frisch}},\ and\ \bibinfo {author} {\bibfnamefont {J.~L.}\ \bibnamefont {Lebowitz}},\ }\bibfield  {title} {\enquote {\bibinfo {title} {Scaled particle theory of fluids},}\ }\href {https://doi.org/10.1063/1.1731417} {\bibfield  {journal} {\bibinfo  {journal} {The Journal of Chemical Physics}\ }\textbf {\bibinfo {volume} {33}},\ \bibinfo {pages} {1379--1385} (\bibinfo {year} {1960})}\BibitemShut {NoStop}%
\bibitem [{\citenamefont {Corti}\ \emph {et~al.}(1997)\citenamefont {Corti}, \citenamefont {Debenedetti}, \citenamefont {Sastry},\ and\ \citenamefont {Stillinger}}]{RN2984}%
  \BibitemOpen
  \bibfield  {author} {\bibinfo {author} {\bibfnamefont {D.~S.}\ \bibnamefont {Corti}}, \bibinfo {author} {\bibfnamefont {P.~G.}\ \bibnamefont {Debenedetti}}, \bibinfo {author} {\bibfnamefont {S.}~\bibnamefont {Sastry}},\ and\ \bibinfo {author} {\bibfnamefont {F.~H.}\ \bibnamefont {Stillinger}},\ }\bibfield  {title} {\enquote {\bibinfo {title} {Constraints, metastability, and inherent structures in liquids},}\ }\href {https://doi.org/10.1103/PhysRevE.55.5522} {\bibfield  {journal} {\bibinfo  {journal} {Physical Review E}\ }\textbf {\bibinfo {volume} {55}},\ \bibinfo {pages} {5522--5534} (\bibinfo {year} {1997})}\BibitemShut {NoStop}%
\bibitem [{\citenamefont {in~‘t Veld}\ \emph {et~al.}(2000)\citenamefont {in~‘t Veld}, \citenamefont {Stone}, \citenamefont {Truskett},\ and\ \citenamefont {Sanchez}}]{RN2869}%
  \BibitemOpen
  \bibfield  {author} {\bibinfo {author} {\bibfnamefont {P.~J.}\ \bibnamefont {in~‘t Veld}}, \bibinfo {author} {\bibfnamefont {M.~T.}\ \bibnamefont {Stone}}, \bibinfo {author} {\bibfnamefont {T.~M.}\ \bibnamefont {Truskett}},\ and\ \bibinfo {author} {\bibfnamefont {I.~C.}\ \bibnamefont {Sanchez}},\ }\bibfield  {title} {\enquote {\bibinfo {title} {Liquid structure via cavity size distributions},}\ }\href {https://doi.org/10.1021/jp001934c} {\bibfield  {journal} {\bibinfo  {journal} {The Journal of Physical Chemistry B}\ }\textbf {\bibinfo {volume} {104}},\ \bibinfo {pages} {12028--12034} (\bibinfo {year} {2000})}\BibitemShut {NoStop}%
\bibitem [{\citenamefont {Simon}\ \emph {et~al.}(2022)\citenamefont {Simon}, \citenamefont {Krüger}, \citenamefont {Schnell}, \citenamefont {Vlugt}, \citenamefont {Kjelstrup},\ and\ \citenamefont {Bedeaux}}]{RN2961}%
  \BibitemOpen
  \bibfield  {author} {\bibinfo {author} {\bibfnamefont {J.~M.}\ \bibnamefont {Simon}}, \bibinfo {author} {\bibfnamefont {P.}~\bibnamefont {Krüger}}, \bibinfo {author} {\bibfnamefont {S.~K.}\ \bibnamefont {Schnell}}, \bibinfo {author} {\bibfnamefont {T.~J.~H.}\ \bibnamefont {Vlugt}}, \bibinfo {author} {\bibfnamefont {S.}~\bibnamefont {Kjelstrup}},\ and\ \bibinfo {author} {\bibfnamefont {D.}~\bibnamefont {Bedeaux}},\ }\bibfield  {title} {\enquote {\bibinfo {title} {{K}irkwood–{B}uff integrals: From fluctuations in finite volumes to the thermodynamic limit},}\ }\href {https://doi.org/10.1063/5.0106162} {\bibfield  {journal} {\bibinfo  {journal} {The Journal of Chemical Physics}\ }\textbf {\bibinfo {volume} {157}},\ \bibinfo {pages} {130901} (\bibinfo {year} {2022})}\BibitemShut {NoStop}%
\bibitem [{\citenamefont {Giri}\ \emph {et~al.}(2015)\citenamefont {Giri}, \citenamefont {Del~Pópolo}, \citenamefont {Melaugh}, \citenamefont {Greenaway}, \citenamefont {Rätzke}, \citenamefont {Koschine}, \citenamefont {Pison}, \citenamefont {Gomes}, \citenamefont {Cooper},\ and\ \citenamefont {James}}]{RN3180}%
  \BibitemOpen
  \bibfield  {author} {\bibinfo {author} {\bibfnamefont {N.}~\bibnamefont {Giri}}, \bibinfo {author} {\bibfnamefont {M.~G.}\ \bibnamefont {Del~Pópolo}}, \bibinfo {author} {\bibfnamefont {G.}~\bibnamefont {Melaugh}}, \bibinfo {author} {\bibfnamefont {R.~L.}\ \bibnamefont {Greenaway}}, \bibinfo {author} {\bibfnamefont {K.}~\bibnamefont {Rätzke}}, \bibinfo {author} {\bibfnamefont {T.}~\bibnamefont {Koschine}}, \bibinfo {author} {\bibfnamefont {L.}~\bibnamefont {Pison}}, \bibinfo {author} {\bibfnamefont {M.~F.~C.}\ \bibnamefont {Gomes}}, \bibinfo {author} {\bibfnamefont {A.~I.}\ \bibnamefont {Cooper}},\ and\ \bibinfo {author} {\bibfnamefont {S.~L.}\ \bibnamefont {James}},\ }\bibfield  {title} {\enquote {\bibinfo {title} {Liquids with permanent porosity},}\ }\href@noop {} {\bibfield  {journal} {\bibinfo  {journal} {Nature}\ }\textbf {\bibinfo {volume} {527}},\ \bibinfo {pages} {216--220} (\bibinfo {year} {2015})}\BibitemShut {NoStop}%
\bibitem [{\citenamefont {Reiss}\ \emph {et~al.}(1960)\citenamefont {Reiss}, \citenamefont {Frisch}, \citenamefont {Helfand},\ and\ \citenamefont {Lebowitz}}]{RN2861}%
  \BibitemOpen
  \bibfield  {author} {\bibinfo {author} {\bibfnamefont {H.}~\bibnamefont {Reiss}}, \bibinfo {author} {\bibfnamefont {H.~L.}\ \bibnamefont {Frisch}}, \bibinfo {author} {\bibfnamefont {E.}~\bibnamefont {Helfand}},\ and\ \bibinfo {author} {\bibfnamefont {J.~L.}\ \bibnamefont {Lebowitz}},\ }\bibfield  {title} {\enquote {\bibinfo {title} {Aspects of the statistical thermodynamics of real fluids},}\ }\href {https://doi.org/10.1063/1.1700883} {\bibfield  {journal} {\bibinfo  {journal} {The Journal of Chemical Physics}\ }\textbf {\bibinfo {volume} {32}},\ \bibinfo {pages} {119--124} (\bibinfo {year} {1960})}\BibitemShut {NoStop}%
\bibitem [{\citenamefont {Postma}, \citenamefont {Berendsen},\ and\ \citenamefont {Haak}(1982)}]{RN2871}%
  \BibitemOpen
  \bibfield  {author} {\bibinfo {author} {\bibfnamefont {J.~P.~M.}\ \bibnamefont {Postma}}, \bibinfo {author} {\bibfnamefont {H.~J.~C.}\ \bibnamefont {Berendsen}},\ and\ \bibinfo {author} {\bibfnamefont {J.~R.}\ \bibnamefont {Haak}},\ }\bibfield  {title} {\enquote {\bibinfo {title} {Thermodynamics of cavity formation in water. {A} molecular dynamics study},}\ }\href {https://doi.org/10.1039/FS9821700055} {\bibfield  {journal} {\bibinfo  {journal} {Faraday Symposia of the Chemical Society}\ }\textbf {\bibinfo {volume} {17}},\ \bibinfo {pages} {55--67} (\bibinfo {year} {1982})}\BibitemShut {NoStop}%
\bibitem [{\citenamefont {Pohorille}\ and\ \citenamefont {Pratt}(1990)}]{RN2983}%
  \BibitemOpen
  \bibfield  {author} {\bibinfo {author} {\bibfnamefont {A.}~\bibnamefont {Pohorille}}\ and\ \bibinfo {author} {\bibfnamefont {L.~R.}\ \bibnamefont {Pratt}},\ }\bibfield  {title} {\enquote {\bibinfo {title} {Cavities in molecular liquids and the theory of hydrophobic solubilities},}\ }\href {https://doi.org/10.1021/ja00169a011} {\bibfield  {journal} {\bibinfo  {journal} {Journal of the American Chemical Society}\ }\textbf {\bibinfo {volume} {112}},\ \bibinfo {pages} {5066--5074} (\bibinfo {year} {1990})}\BibitemShut {NoStop}%
\bibitem [{\citenamefont {Arvengas}\ \emph {et~al.}(2011)\citenamefont {Arvengas}, \citenamefont {Herbert}, \citenamefont {Cersoy}, \citenamefont {Davitt},\ and\ \citenamefont {Caupin}}]{RN2997}%
  \BibitemOpen
  \bibfield  {author} {\bibinfo {author} {\bibfnamefont {A.}~\bibnamefont {Arvengas}}, \bibinfo {author} {\bibfnamefont {E.}~\bibnamefont {Herbert}}, \bibinfo {author} {\bibfnamefont {S.}~\bibnamefont {Cersoy}}, \bibinfo {author} {\bibfnamefont {K.}~\bibnamefont {Davitt}},\ and\ \bibinfo {author} {\bibfnamefont {F.}~\bibnamefont {Caupin}},\ }\bibfield  {title} {\enquote {\bibinfo {title} {Cavitation in heavy water and other liquids},}\ }\href {https://doi.org/10.1021/jp2050977} {\bibfield  {journal} {\bibinfo  {journal} {The Journal of Physical Chemistry B}\ }\textbf {\bibinfo {volume} {115}},\ \bibinfo {pages} {14240--14245} (\bibinfo {year} {2011})}\BibitemShut {NoStop}%
\bibitem [{\citenamefont {Baidakov}\ and\ \citenamefont {Bobrov}(2014)}]{RN2760}%
  \BibitemOpen
  \bibfield  {author} {\bibinfo {author} {\bibfnamefont {V.~G.}\ \bibnamefont {Baidakov}}\ and\ \bibinfo {author} {\bibfnamefont {K.~S.}\ \bibnamefont {Bobrov}},\ }\bibfield  {title} {\enquote {\bibinfo {title} {Spontaneous cavitation in a {L}ennard-{J}ones liquid at negative pressures},}\ }\href {https://doi.org/10.1063/1.4874644} {\bibfield  {journal} {\bibinfo  {journal} {The Journal of Chemical Physics}\ }\textbf {\bibinfo {volume} {140}},\ \bibinfo {pages} {184506} (\bibinfo {year} {2014})}\BibitemShut {NoStop}%
\bibitem [{\citenamefont {Baidakov}\ and\ \citenamefont {Protsenko}(2020)}]{RN2758}%
  \BibitemOpen
  \bibfield  {author} {\bibinfo {author} {\bibfnamefont {V.~G.}\ \bibnamefont {Baidakov}}\ and\ \bibinfo {author} {\bibfnamefont {K.~R.}\ \bibnamefont {Protsenko}},\ }\bibfield  {title} {\enquote {\bibinfo {title} {Molecular dynamics simulation of cavitation in a {L}ennard-{J}ones liquid at negative pressures},}\ }\href {https://doi.org/10.1016/j.cplett.2020.138030} {\bibfield  {journal} {\bibinfo  {journal} {Chemical Physics Letters}\ }\textbf {\bibinfo {volume} {760}},\ \bibinfo {pages} {138030} (\bibinfo {year} {2020})}\BibitemShut {NoStop}%
\bibitem [{\citenamefont {Doebele}\ \emph {et~al.}(2020)\citenamefont {Doebele}, \citenamefont {Benoit-Gonin}, \citenamefont {Souris}, \citenamefont {Cagnon}, \citenamefont {Spathis}, \citenamefont {Wolf}, \citenamefont {Grosman}, \citenamefont {Bossert}, \citenamefont {Trimaille},\ and\ \citenamefont {Rolley}}]{RN2715}%
  \BibitemOpen
  \bibfield  {author} {\bibinfo {author} {\bibfnamefont {V.}~\bibnamefont {Doebele}}, \bibinfo {author} {\bibfnamefont {A.}~\bibnamefont {Benoit-Gonin}}, \bibinfo {author} {\bibfnamefont {F.}~\bibnamefont {Souris}}, \bibinfo {author} {\bibfnamefont {L.}~\bibnamefont {Cagnon}}, \bibinfo {author} {\bibfnamefont {P.}~\bibnamefont {Spathis}}, \bibinfo {author} {\bibfnamefont {P.-E.}\ \bibnamefont {Wolf}}, \bibinfo {author} {\bibfnamefont {A.}~\bibnamefont {Grosman}}, \bibinfo {author} {\bibfnamefont {M.}~\bibnamefont {Bossert}}, \bibinfo {author} {\bibfnamefont {I.}~\bibnamefont {Trimaille}},\ and\ \bibinfo {author} {\bibfnamefont {E.}~\bibnamefont {Rolley}},\ }\bibfield  {title} {\enquote {\bibinfo {title} {Direct observation of homogeneous cavitation in nanopores},}\ }\href {https://doi.org/10.1103/PhysRevLett.125.255701} {\bibfield  {journal} {\bibinfo  {journal} {Physical Review Letters}\ }\textbf {\bibinfo {volume} {125}},\ \bibinfo {pages} {255701} (\bibinfo {year} {2020})}\BibitemShut {NoStop}%
\bibitem [{\citenamefont {Puibasset}(2021)}]{RN2845}%
  \BibitemOpen
  \bibfield  {author} {\bibinfo {author} {\bibfnamefont {J.}~\bibnamefont {Puibasset}},\ }\bibfield  {title} {\enquote {\bibinfo {title} {Cavitation in heterogeneous nanopores: The chemical ink-bottle},}\ }\href {https://doi.org/10.1063/5.0065166} {\bibfield  {journal} {\bibinfo  {journal} {AIP Advances}\ }\textbf {\bibinfo {volume} {11}},\ \bibinfo {pages} {095311} (\bibinfo {year} {2021})}\BibitemShut {NoStop}%
\bibitem [{\citenamefont {Bossert}\ \emph {et~al.}(2021)\citenamefont {Bossert}, \citenamefont {Grosman}, \citenamefont {Trimaille}, \citenamefont {Souris}, \citenamefont {Doebele}, \citenamefont {Benoit-Gonin}, \citenamefont {Cagnon}, \citenamefont {Spathis}, \citenamefont {Wolf},\ and\ \citenamefont {Rolley}}]{RN2972}%
  \BibitemOpen
  \bibfield  {author} {\bibinfo {author} {\bibfnamefont {M.}~\bibnamefont {Bossert}}, \bibinfo {author} {\bibfnamefont {A.}~\bibnamefont {Grosman}}, \bibinfo {author} {\bibfnamefont {I.}~\bibnamefont {Trimaille}}, \bibinfo {author} {\bibfnamefont {F.}~\bibnamefont {Souris}}, \bibinfo {author} {\bibfnamefont {V.}~\bibnamefont {Doebele}}, \bibinfo {author} {\bibfnamefont {A.}~\bibnamefont {Benoit-Gonin}}, \bibinfo {author} {\bibfnamefont {L.}~\bibnamefont {Cagnon}}, \bibinfo {author} {\bibfnamefont {P.}~\bibnamefont {Spathis}}, \bibinfo {author} {\bibfnamefont {P.-E.}\ \bibnamefont {Wolf}},\ and\ \bibinfo {author} {\bibfnamefont {E.}~\bibnamefont {Rolley}},\ }\bibfield  {title} {\enquote {\bibinfo {title} {Evaporation process in porous silicon: Cavitation vs pore blocking},}\ }\href {https://doi.org/10.1021/acs.langmuir.1c02397} {\bibfield  {journal} {\bibinfo  {journal} {Langmuir}\ }\textbf {\bibinfo {volume} {37}},\ \bibinfo {pages} {14419--14428} (\bibinfo {year} {2021})}\BibitemShut {NoStop}%
\bibitem [{\citenamefont {Bossert}\ \emph {et~al.}(2023)\citenamefont {Bossert}, \citenamefont {Trimaille}, \citenamefont {Cagnon}, \citenamefont {Chabaud}, \citenamefont {Gueneau}, \citenamefont {Spathis}, \citenamefont {Wolf},\ and\ \citenamefont {Rolley}}]{RN2996}%
  \BibitemOpen
  \bibfield  {author} {\bibinfo {author} {\bibfnamefont {M.}~\bibnamefont {Bossert}}, \bibinfo {author} {\bibfnamefont {I.}~\bibnamefont {Trimaille}}, \bibinfo {author} {\bibfnamefont {L.}~\bibnamefont {Cagnon}}, \bibinfo {author} {\bibfnamefont {B.}~\bibnamefont {Chabaud}}, \bibinfo {author} {\bibfnamefont {C.}~\bibnamefont {Gueneau}}, \bibinfo {author} {\bibfnamefont {P.}~\bibnamefont {Spathis}}, \bibinfo {author} {\bibfnamefont {P.~E.}\ \bibnamefont {Wolf}},\ and\ \bibinfo {author} {\bibfnamefont {E.}~\bibnamefont {Rolley}},\ }\bibfield  {title} {\enquote {\bibinfo {title} {Surface tension of cavitation bubbles},}\ }\href {https://doi.org/10.1073/pnas.2300499120} {\bibfield  {journal} {\bibinfo  {journal} {Proceedings of the National Academy of Sciences of the United States of America}\ }\textbf {\bibinfo {volume} {120}},\ \bibinfo {pages} {e2300499120} (\bibinfo {year} {2023})}\BibitemShut {NoStop}%
\bibitem [{\citenamefont {Bal}\ and\ \citenamefont {Neyts}(2022)}]{RN2991}%
  \BibitemOpen
  \bibfield  {author} {\bibinfo {author} {\bibfnamefont {K.~M.}\ \bibnamefont {Bal}}\ and\ \bibinfo {author} {\bibfnamefont {E.~C.}\ \bibnamefont {Neyts}},\ }\bibfield  {title} {\enquote {\bibinfo {title} {Extending and validating bubble nucleation rate predictions in a {L}ennard-{J}ones fluid with enhanced sampling methods and transition state theory},}\ }\href {https://doi.org/10.1063/5.0120136} {\bibfield  {journal} {\bibinfo  {journal} {The Journal of Chemical Physics}\ }\textbf {\bibinfo {volume} {157}},\ \bibinfo {pages} {184113} (\bibinfo {year} {2022})}\BibitemShut {NoStop}%
\bibitem [{\citenamefont {Protsenko}\ and\ \citenamefont {Baidakov}(2023)}]{RN3000}%
  \BibitemOpen
  \bibfield  {author} {\bibinfo {author} {\bibfnamefont {K.~R.}\ \bibnamefont {Protsenko}}\ and\ \bibinfo {author} {\bibfnamefont {V.~G.}\ \bibnamefont {Baidakov}},\ }\bibfield  {title} {\enquote {\bibinfo {title} {Classical nucleation theory and molecular dynamics simulation: {C}avitation},}\ }\href {https://doi.org/10.1063/5.0134778} {\bibfield  {journal} {\bibinfo  {journal} {Physics of Fluids}\ }\textbf {\bibinfo {volume} {35}},\ \bibinfo {pages} {014111} (\bibinfo {year} {2023})}\BibitemShut {NoStop}%
\bibitem [{\citenamefont {Lamas}\ \emph {et~al.}(2023)\citenamefont {Lamas}, \citenamefont {Sanz}, \citenamefont {Vega},\ and\ \citenamefont {Noya}}]{RN2989}%
  \BibitemOpen
  \bibfield  {author} {\bibinfo {author} {\bibfnamefont {C.~P.}\ \bibnamefont {Lamas}}, \bibinfo {author} {\bibfnamefont {E.}~\bibnamefont {Sanz}}, \bibinfo {author} {\bibfnamefont {C.}~\bibnamefont {Vega}},\ and\ \bibinfo {author} {\bibfnamefont {E.~G.}\ \bibnamefont {Noya}},\ }\bibfield  {title} {\enquote {\bibinfo {title} {Estimation of bubble cavitation rates in a symmetrical {L}ennard-{J}ones mixture by {NVT} seeding simulations},}\ }\href {https://doi.org/10.1063/5.0142109} {\bibfield  {journal} {\bibinfo  {journal} {The Journal of Chemical Physics}\ }\textbf {\bibinfo {volume} {158}},\ \bibinfo {pages} {124109} (\bibinfo {year} {2023})}\BibitemShut {NoStop}%
\bibitem [{\citenamefont {Fisher}(1948)}]{RN548}%
  \BibitemOpen
  \bibfield  {author} {\bibinfo {author} {\bibfnamefont {J.~C.}\ \bibnamefont {Fisher}},\ }\bibfield  {title} {\enquote {\bibinfo {title} {The fracture of liquids},}\ }\href {https://doi.org/10.1063/1.1698012} {\bibfield  {journal} {\bibinfo  {journal} {Journal of Applied Physics}\ }\textbf {\bibinfo {volume} {19}},\ \bibinfo {pages} {1062--1067} (\bibinfo {year} {1948})}\BibitemShut {NoStop}%
\bibitem [{\citenamefont {Blander}\ and\ \citenamefont {Katz}(1975)}]{RN512}%
  \BibitemOpen
  \bibfield  {author} {\bibinfo {author} {\bibfnamefont {M.}~\bibnamefont {Blander}}\ and\ \bibinfo {author} {\bibfnamefont {J.~L.}\ \bibnamefont {Katz}},\ }\bibfield  {title} {\enquote {\bibinfo {title} {Bubble nucleation in liquids},}\ }\href {https://doi.org/10.1002/aic.690210502} {\bibfield  {journal} {\bibinfo  {journal} {American Institute of Chemical Engineers Journal}\ }\textbf {\bibinfo {volume} {21}},\ \bibinfo {pages} {833--848} (\bibinfo {year} {1975})}\BibitemShut {NoStop}%
\bibitem [{\citenamefont {Debenedetti}(1996)}]{RN2920}%
  \BibitemOpen
  \bibfield  {author} {\bibinfo {author} {\bibfnamefont {P.~G.}\ \bibnamefont {Debenedetti}},\ }\href@noop {} {\emph {\bibinfo {title} {Metastable Liquids: Concepts and Principles}}}\ (\bibinfo  {publisher} {Princeton University Press},\ \bibinfo {address} {Princeton, NJ},\ \bibinfo {year} {1996})\BibitemShut {NoStop}%
\bibitem [{\citenamefont {Gibbs}(1906)}]{RN2076}%
  \BibitemOpen
  \bibfield  {author} {\bibinfo {author} {\bibfnamefont {J.~W.}\ \bibnamefont {Gibbs}},\ }\href@noop {} {\emph {\bibinfo {title} {The Scientific Papers of J Willard Gibbs}}}\ (\bibinfo  {publisher} {Longmans Green},\ \bibinfo {address} {London},\ \bibinfo {year} {1906})\BibitemShut {NoStop}%
\bibitem [{\citenamefont {Reiss}\ and\ \citenamefont {Bowles}(1999)}]{RN2913}%
  \BibitemOpen
  \bibfield  {author} {\bibinfo {author} {\bibfnamefont {H.}~\bibnamefont {Reiss}}\ and\ \bibinfo {author} {\bibfnamefont {R.~K.}\ \bibnamefont {Bowles}},\ }\bibfield  {title} {\enquote {\bibinfo {title} {Some fundamental statistical mechanical relations concerning physical clusters of interest to nucleation theory},}\ }\href {https://doi.org/10.1063/1.480075} {\bibfield  {journal} {\bibinfo  {journal} {The Journal of Chemical Physics}\ }\textbf {\bibinfo {volume} {111}},\ \bibinfo {pages} {7501--7504} (\bibinfo {year} {1999})}\BibitemShut {NoStop}%
\bibitem [{\citenamefont {Kinjo}\ and\ \citenamefont {Matsumoto}(1998)}]{RN2756}%
  \BibitemOpen
  \bibfield  {author} {\bibinfo {author} {\bibfnamefont {T.}~\bibnamefont {Kinjo}}\ and\ \bibinfo {author} {\bibfnamefont {M.}~\bibnamefont {Matsumoto}},\ }\bibfield  {title} {\enquote {\bibinfo {title} {Cavitation processes and negative pressure},}\ }\href {https://doi.org/10.1016/S0378-3812(97)00278-1} {\bibfield  {journal} {\bibinfo  {journal} {Fluid Phase Equilibria}\ }\textbf {\bibinfo {volume} {144}},\ \bibinfo {pages} {343--350} (\bibinfo {year} {1998})}\BibitemShut {NoStop}%
\bibitem [{\citenamefont {Menzl}\ \emph {et~al.}(2016)\citenamefont {Menzl}, \citenamefont {Gonzalez}, \citenamefont {Geiger}, \citenamefont {Caupin}, \citenamefont {Abascal}, \citenamefont {Valeriani},\ and\ \citenamefont {Dellago}}]{RN2801}%
  \BibitemOpen
  \bibfield  {author} {\bibinfo {author} {\bibfnamefont {G.}~\bibnamefont {Menzl}}, \bibinfo {author} {\bibfnamefont {M.~A.}\ \bibnamefont {Gonzalez}}, \bibinfo {author} {\bibfnamefont {P.}~\bibnamefont {Geiger}}, \bibinfo {author} {\bibfnamefont {F.}~\bibnamefont {Caupin}}, \bibinfo {author} {\bibfnamefont {J.~L.~F.}\ \bibnamefont {Abascal}}, \bibinfo {author} {\bibfnamefont {C.}~\bibnamefont {Valeriani}},\ and\ \bibinfo {author} {\bibfnamefont {C.}~\bibnamefont {Dellago}},\ }\bibfield  {title} {\enquote {\bibinfo {title} {Molecular mechanism for cavitation in water under tension},}\ }\href {https://doi.org/10.1073/pnas.1608421113} {\bibfield  {journal} {\bibinfo  {journal} {Proceedings of the National Academy of Sciences USA}\ }\textbf {\bibinfo {volume} {113}},\ \bibinfo {pages} {13582} (\bibinfo {year} {2016})}\BibitemShut {NoStop}%
\bibitem [{\citenamefont {Shen}\ and\ \citenamefont {Debenedetti}(1999)}]{RN2968}%
  \BibitemOpen
  \bibfield  {author} {\bibinfo {author} {\bibfnamefont {V.~K.}\ \bibnamefont {Shen}}\ and\ \bibinfo {author} {\bibfnamefont {P.~G.}\ \bibnamefont {Debenedetti}},\ }\bibfield  {title} {\enquote {\bibinfo {title} {A computational study of homogeneous liquid–vapor nucleation in the {L}ennard-{J}ones fluid},}\ }\href {https://doi.org/10.1063/1.479639} {\bibfield  {journal} {\bibinfo  {journal} {The Journal of Chemical Physics}\ }\textbf {\bibinfo {volume} {111}},\ \bibinfo {pages} {3581--3589} (\bibinfo {year} {1999})}\BibitemShut {NoStop}%
\bibitem [{\citenamefont {Vishnyakov}, \citenamefont {Debenedetti},\ and\ \citenamefont {Neimark}(2000)}]{RN2985}%
  \BibitemOpen
  \bibfield  {author} {\bibinfo {author} {\bibfnamefont {A.}~\bibnamefont {Vishnyakov}}, \bibinfo {author} {\bibfnamefont {P.~G.}\ \bibnamefont {Debenedetti}},\ and\ \bibinfo {author} {\bibfnamefont {A.~V.}\ \bibnamefont {Neimark}},\ }\bibfield  {title} {\enquote {\bibinfo {title} {Statistical geometry of cavities in a metastable confined fluid},}\ }\href {https://doi.org/10.1103/PhysRevE.62.538} {\bibfield  {journal} {\bibinfo  {journal} {Physical Review E}\ }\textbf {\bibinfo {volume} {62}},\ \bibinfo {pages} {538--544} (\bibinfo {year} {2000})}\BibitemShut {NoStop}%
\bibitem [{\citenamefont {Wu}\ and\ \citenamefont {Pan}(2003)}]{RN2770}%
  \BibitemOpen
  \bibfield  {author} {\bibinfo {author} {\bibfnamefont {Y.~W.}\ \bibnamefont {Wu}}\ and\ \bibinfo {author} {\bibfnamefont {C.}~\bibnamefont {Pan}},\ }\bibfield  {title} {\enquote {\bibinfo {title} {A molecular dynamics simulation of bubble nucleation in homogeneous liquid under heating with constant mean negative pressure},}\ }\href {https://doi.org/10.1080/10893950390203323} {\bibfield  {journal} {\bibinfo  {journal} {Microscale Thermophysical Engineering}\ }\textbf {\bibinfo {volume} {7}},\ \bibinfo {pages} {137--151} (\bibinfo {year} {2003})}\BibitemShut {NoStop}%
\bibitem [{\citenamefont {Neimark}\ and\ \citenamefont {Vishnyakov}(2005)}]{RN598}%
  \BibitemOpen
  \bibfield  {author} {\bibinfo {author} {\bibfnamefont {A.~V.}\ \bibnamefont {Neimark}}\ and\ \bibinfo {author} {\bibfnamefont {A.}~\bibnamefont {Vishnyakov}},\ }\bibfield  {title} {\enquote {\bibinfo {title} {The birth of a bubble: a molecular simulation study},}\ }\href {https://doi.org/10.1063/1.1829040} {\bibfield  {journal} {\bibinfo  {journal} {The Journal of Chemical Physics}\ }\textbf {\bibinfo {volume} {122}},\ \bibinfo {pages} {054707} (\bibinfo {year} {2005})}\BibitemShut {NoStop}%
\bibitem [{\citenamefont {Wedekind}\ \emph {et~al.}(2009)\citenamefont {Wedekind}, \citenamefont {Chkonia}, \citenamefont {Wölk}, \citenamefont {Strey},\ and\ \citenamefont {Reguera}}]{RN2903}%
  \BibitemOpen
  \bibfield  {author} {\bibinfo {author} {\bibfnamefont {J.}~\bibnamefont {Wedekind}}, \bibinfo {author} {\bibfnamefont {G.}~\bibnamefont {Chkonia}}, \bibinfo {author} {\bibfnamefont {J.}~\bibnamefont {Wölk}}, \bibinfo {author} {\bibfnamefont {R.}~\bibnamefont {Strey}},\ and\ \bibinfo {author} {\bibfnamefont {D.}~\bibnamefont {Reguera}},\ }\bibfield  {title} {\enquote {\bibinfo {title} {Crossover from nucleation to spinodal decomposition in a condensing vapor},}\ }\href {https://doi.org/10.1063/1.3204448} {\bibfield  {journal} {\bibinfo  {journal} {The Journal of Chemical Physics}\ }\textbf {\bibinfo {volume} {131}},\ \bibinfo {pages} {114506} (\bibinfo {year} {2009})}\BibitemShut {NoStop}%
\bibitem [{\citenamefont {Watanabe}, \citenamefont {Suzuki},\ and\ \citenamefont {Ito}(2010)}]{RN2766}%
  \BibitemOpen
  \bibfield  {author} {\bibinfo {author} {\bibfnamefont {H.}~\bibnamefont {Watanabe}}, \bibinfo {author} {\bibfnamefont {M.}~\bibnamefont {Suzuki}},\ and\ \bibinfo {author} {\bibfnamefont {N.}~\bibnamefont {Ito}},\ }\bibfield  {title} {\enquote {\bibinfo {title} {Cumulative distribution functions associated with bubble-nucleation processes in cavitation},}\ }\href {https://doi.org/10.1103/PhysRevE.82.051604} {\bibfield  {journal} {\bibinfo  {journal} {Physical Review E}\ }\textbf {\bibinfo {volume} {82}},\ \bibinfo {pages} {051604} (\bibinfo {year} {2010})}\BibitemShut {NoStop}%
\bibitem [{\citenamefont {Meadley}\ and\ \citenamefont {Escobedo}(2012)}]{RN2797}%
  \BibitemOpen
  \bibfield  {author} {\bibinfo {author} {\bibfnamefont {S.~L.}\ \bibnamefont {Meadley}}\ and\ \bibinfo {author} {\bibfnamefont {F.~A.}\ \bibnamefont {Escobedo}},\ }\bibfield  {title} {\enquote {\bibinfo {title} {Thermodynamics and kinetics of bubble nucleation: Simulation methodology},}\ }\href {https://doi.org/10.1063/1.4745082} {\bibfield  {journal} {\bibinfo  {journal} {The Journal of Chemical Physics}\ }\textbf {\bibinfo {volume} {137}},\ \bibinfo {pages} {074109} (\bibinfo {year} {2012})}\BibitemShut {NoStop}%
\bibitem [{\citenamefont {Frenkel}(1939{\natexlab{a}})}]{RN3177}%
  \BibitemOpen
  \bibfield  {author} {\bibinfo {author} {\bibfnamefont {J.}~\bibnamefont {Frenkel}},\ }\bibfield  {title} {\enquote {\bibinfo {title} {A general theory of heterophase fluctuations and pretransition phenomena},}\ }\href@noop {} {\bibfield  {journal} {\bibinfo  {journal} {The Journal of Chemical Physics}\ }\textbf {\bibinfo {volume} {7}},\ \bibinfo {pages} {538--547} (\bibinfo {year} {1939}{\natexlab{a}})}\BibitemShut {NoStop}%
\bibitem [{\citenamefont {Frenkel}(1939{\natexlab{b}})}]{RN3141}%
  \BibitemOpen
  \bibfield  {author} {\bibinfo {author} {\bibfnamefont {J.}~\bibnamefont {Frenkel}},\ }\bibfield  {title} {\enquote {\bibinfo {title} {Statistical theory of condensation phenomena},}\ }\href@noop {} {\bibfield  {journal} {\bibinfo  {journal} {The Journal of Chemical Physics}\ }\textbf {\bibinfo {volume} {7}},\ \bibinfo {pages} {200--201} (\bibinfo {year} {1939}{\natexlab{b}})}\BibitemShut {NoStop}%
\bibitem [{\citenamefont {Oh}\ and\ \citenamefont {Zeng}(1999)}]{RN3145}%
  \BibitemOpen
  \bibfield  {author} {\bibinfo {author} {\bibfnamefont {K.~J.}\ \bibnamefont {Oh}}\ and\ \bibinfo {author} {\bibfnamefont {X.~C.}\ \bibnamefont {Zeng}},\ }\bibfield  {title} {\enquote {\bibinfo {title} {Formation free energy of clusters in vapor-liquid nucleation: A {M}onte {C}arlo simulation study},}\ }\href@noop {} {\bibfield  {journal} {\bibinfo  {journal} {The Journal of Chemical Physics}\ }\textbf {\bibinfo {volume} {110}},\ \bibinfo {pages} {4471--4476} (\bibinfo {year} {1999})}\BibitemShut {NoStop}%
\bibitem [{\citenamefont {ten Wolde}\ and\ \citenamefont {Frenkel}(1998)}]{RN1956}%
  \BibitemOpen
  \bibfield  {author} {\bibinfo {author} {\bibfnamefont {P.~R.}\ \bibnamefont {ten Wolde}}\ and\ \bibinfo {author} {\bibfnamefont {D.}~\bibnamefont {Frenkel}},\ }\bibfield  {title} {\enquote {\bibinfo {title} {Computer simulation study of gas-liquid nucleation in a {L}ennard-{J}ones system},}\ }\href {https://doi.org/10.1063/1.477658} {\bibfield  {journal} {\bibinfo  {journal} {The Journal of Chemical Physics}\ }\textbf {\bibinfo {volume} {109}},\ \bibinfo {pages} {9901--9918} (\bibinfo {year} {1998})}\BibitemShut {NoStop}%
\bibitem [{\citenamefont {Kusaka}, \citenamefont {Wang},\ and\ \citenamefont {Seinfeld}(1998)}]{RN3140}%
  \BibitemOpen
  \bibfield  {author} {\bibinfo {author} {\bibfnamefont {I.}~\bibnamefont {Kusaka}}, \bibinfo {author} {\bibfnamefont {Z.~G.}\ \bibnamefont {Wang}},\ and\ \bibinfo {author} {\bibfnamefont {J.~H.}\ \bibnamefont {Seinfeld}},\ }\bibfield  {title} {\enquote {\bibinfo {title} {Direct evaluation of the equilibrium distribution of physical clusters by a grand canonical {M}onte {C}arlo simulation},}\ }\href@noop {} {\bibfield  {journal} {\bibinfo  {journal} {The Journal of Chemical Physics}\ }\textbf {\bibinfo {volume} {108}},\ \bibinfo {pages} {3416--3423} (\bibinfo {year} {1998})}\BibitemShut {NoStop}%
\bibitem [{\citenamefont {Yasuoka}\ and\ \citenamefont {Matsumoto}(1998{\natexlab{a}})}]{RN2932}%
  \BibitemOpen
  \bibfield  {author} {\bibinfo {author} {\bibfnamefont {K.}~\bibnamefont {Yasuoka}}\ and\ \bibinfo {author} {\bibfnamefont {M.}~\bibnamefont {Matsumoto}},\ }\bibfield  {title} {\enquote {\bibinfo {title} {Molecular dynamics of homogeneous nucleation in the vapor phase. {I}. {L}ennard-{J}ones fluid},}\ }\href@noop {} {\bibfield  {journal} {\bibinfo  {journal} {The Journal of Chemical Physics}\ }\textbf {\bibinfo {volume} {109}},\ \bibinfo {pages} {8451--8462} (\bibinfo {year} {1998}{\natexlab{a}})}\BibitemShut {NoStop}%
\bibitem [{\citenamefont {Yasuoka}\ and\ \citenamefont {Matsumoto}(1998{\natexlab{b}})}]{RN3139}%
  \BibitemOpen
  \bibfield  {author} {\bibinfo {author} {\bibfnamefont {K.}~\bibnamefont {Yasuoka}}\ and\ \bibinfo {author} {\bibfnamefont {M.}~\bibnamefont {Matsumoto}},\ }\bibfield  {title} {\enquote {\bibinfo {title} {Molecular dynamics of homogeneous nucleation in the vapor phase. {II}. {W}ater},}\ }\href@noop {} {\bibfield  {journal} {\bibinfo  {journal} {The Journal of Chemical Physics}\ }\textbf {\bibinfo {volume} {109}},\ \bibinfo {pages} {8463--8470} (\bibinfo {year} {1998}{\natexlab{b}})}\BibitemShut {NoStop}%
\bibitem [{\citenamefont {Auer}\ and\ \citenamefont {Frenkel}(2004)}]{RN2911}%
  \BibitemOpen
  \bibfield  {author} {\bibinfo {author} {\bibfnamefont {S.}~\bibnamefont {Auer}}\ and\ \bibinfo {author} {\bibfnamefont {D.}~\bibnamefont {Frenkel}},\ }\bibfield  {title} {\enquote {\bibinfo {title} {Numerical prediction of absolute crystallization rates in hard-sphere colloids},}\ }\href {https://doi.org/10.1063/1.1638740} {\bibfield  {journal} {\bibinfo  {journal} {The Journal of Chemical Physics}\ }\textbf {\bibinfo {volume} {120}},\ \bibinfo {pages} {3015--3029} (\bibinfo {year} {2004})}\BibitemShut {NoStop}%
\bibitem [{\citenamefont {Ellerby}, \citenamefont {Weakliem},\ and\ \citenamefont {Reiss}(1991)}]{RN3172}%
  \BibitemOpen
  \bibfield  {author} {\bibinfo {author} {\bibfnamefont {H.~M.}\ \bibnamefont {Ellerby}}, \bibinfo {author} {\bibfnamefont {C.~L.}\ \bibnamefont {Weakliem}},\ and\ \bibinfo {author} {\bibfnamefont {H.}~\bibnamefont {Reiss}},\ }\bibfield  {title} {\enquote {\bibinfo {title} {Toward a molecular theory of vapor‐phase nucleation. {I}. {I}dentification of the average embryo},}\ }\href@noop {} {\bibfield  {journal} {\bibinfo  {journal} {The Journal of Chemical Physics}\ }\textbf {\bibinfo {volume} {95}},\ \bibinfo {pages} {9209--9218} (\bibinfo {year} {1991})}\BibitemShut {NoStop}%
\bibitem [{\citenamefont {Ellerby}\ and\ \citenamefont {Reiss}(1992)}]{RN2966}%
  \BibitemOpen
  \bibfield  {author} {\bibinfo {author} {\bibfnamefont {H.~M.}\ \bibnamefont {Ellerby}}\ and\ \bibinfo {author} {\bibfnamefont {H.}~\bibnamefont {Reiss}},\ }\bibfield  {title} {\enquote {\bibinfo {title} {Toward a molecular theory of vapor‐phase nucleation. {II}. {F}undamental treatment of the cluster distribution},}\ }\href {https://doi.org/10.1063/1.463760} {\bibfield  {journal} {\bibinfo  {journal} {The Journal of Chemical Physics}\ }\textbf {\bibinfo {volume} {97}},\ \bibinfo {pages} {5766--5772} (\bibinfo {year} {1992})}\BibitemShut {NoStop}%
\bibitem [{\citenamefont {ten Wolde}, \citenamefont {Ruiz-Montero},\ and\ \citenamefont {Frenkel}(1996)}]{RN2901}%
  \BibitemOpen
  \bibfield  {author} {\bibinfo {author} {\bibfnamefont {P.-R.}\ \bibnamefont {ten Wolde}}, \bibinfo {author} {\bibfnamefont {M.~J.}\ \bibnamefont {Ruiz-Montero}},\ and\ \bibinfo {author} {\bibfnamefont {D.}~\bibnamefont {Frenkel}},\ }\bibfield  {title} {\enquote {\bibinfo {title} {Simulation of homogeneous crystal nucleation close to coexistence},}\ }\href {https://doi.org/10.1039/FD9960400093} {\bibfield  {journal} {\bibinfo  {journal} {Faraday Discussions}\ }\textbf {\bibinfo {volume} {104}},\ \bibinfo {pages} {93--110} (\bibinfo {year} {1996})}\BibitemShut {NoStop}%
\bibitem [{\citenamefont {Pan}\ and\ \citenamefont {Chandler}(2004)}]{RN2915}%
  \BibitemOpen
  \bibfield  {author} {\bibinfo {author} {\bibfnamefont {A.~C.}\ \bibnamefont {Pan}}\ and\ \bibinfo {author} {\bibfnamefont {D.}~\bibnamefont {Chandler}},\ }\bibfield  {title} {\enquote {\bibinfo {title} {Dynamics of nucleation in the {I}sing model},}\ }\href {https://doi.org/10.1021/jp0471249} {\bibfield  {journal} {\bibinfo  {journal} {J. Phys. Chem. B}\ }\textbf {\bibinfo {volume} {108}},\ \bibinfo {pages} {19681--19686} (\bibinfo {year} {2004})}\BibitemShut {NoStop}%
\bibitem [{\citenamefont {Saika-Voivod}, \citenamefont {Poole},\ and\ \citenamefont {Bowles}(2006)}]{RN2905}%
  \BibitemOpen
  \bibfield  {author} {\bibinfo {author} {\bibfnamefont {I.}~\bibnamefont {Saika-Voivod}}, \bibinfo {author} {\bibfnamefont {P.~H.}\ \bibnamefont {Poole}},\ and\ \bibinfo {author} {\bibfnamefont {R.~K.}\ \bibnamefont {Bowles}},\ }\bibfield  {title} {\enquote {\bibinfo {title} {Test of classical nucleation theory on deeply supercooled high-pressure simulated silica},}\ }\href {https://doi.org/10.1063/1.2203631} {\bibfield  {journal} {\bibinfo  {journal} {The Journal of Chemical Physics}\ }\textbf {\bibinfo {volume} {124}},\ \bibinfo {pages} {224709} (\bibinfo {year} {2006})}\BibitemShut {NoStop}%
\bibitem [{\citenamefont {Lundrigan}\ and\ \citenamefont {Saika-Voivod}(2009)}]{RN2907}%
  \BibitemOpen
  \bibfield  {author} {\bibinfo {author} {\bibfnamefont {S.~E.~M.}\ \bibnamefont {Lundrigan}}\ and\ \bibinfo {author} {\bibfnamefont {I.}~\bibnamefont {Saika-Voivod}},\ }\bibfield  {title} {\enquote {\bibinfo {title} {Test of classical nucleation theory and mean first-passage time formalism on crystallization in the {L}ennard-{J}ones liquid},}\ }\href {https://doi.org/10.1063/1.3216867} {\bibfield  {journal} {\bibinfo  {journal} {The Journal of Chemical Physics}\ }\textbf {\bibinfo {volume} {131}},\ \bibinfo {pages} {104503} (\bibinfo {year} {2009})}\BibitemShut {NoStop}%
\bibitem [{\citenamefont {Saika-Voivod}, \citenamefont {Bowles},\ and\ \citenamefont {Poole}(2009)}]{RN2923}%
  \BibitemOpen
  \bibfield  {author} {\bibinfo {author} {\bibfnamefont {I.}~\bibnamefont {Saika-Voivod}}, \bibinfo {author} {\bibfnamefont {R.~K.}\ \bibnamefont {Bowles}},\ and\ \bibinfo {author} {\bibfnamefont {P.~H.}\ \bibnamefont {Poole}},\ }\bibfield  {title} {\enquote {\bibinfo {title} {Crystal nucleation in a supercooled liquid with glassy dynamics},}\ }\href {https://doi.org/10.1103/PhysRevLett.103.225701} {\bibfield  {journal} {\bibinfo  {journal} {Physical Review Letters}\ }\textbf {\bibinfo {volume} {103}},\ \bibinfo {pages} {225701} (\bibinfo {year} {2009})}\BibitemShut {NoStop}%
\bibitem [{\citenamefont {Goswami}\ \emph {et~al.}(2021)\citenamefont {Goswami}, \citenamefont {Vasisht}, \citenamefont {Frenkel}, \citenamefont {Debenedetti},\ and\ \citenamefont {Sastry}}]{RN2899}%
  \BibitemOpen
  \bibfield  {author} {\bibinfo {author} {\bibfnamefont {Y.}~\bibnamefont {Goswami}}, \bibinfo {author} {\bibfnamefont {V.~V.}\ \bibnamefont {Vasisht}}, \bibinfo {author} {\bibfnamefont {D.}~\bibnamefont {Frenkel}}, \bibinfo {author} {\bibfnamefont {P.~G.}\ \bibnamefont {Debenedetti}},\ and\ \bibinfo {author} {\bibfnamefont {S.}~\bibnamefont {Sastry}},\ }\bibfield  {title} {\enquote {\bibinfo {title} {Thermodynamics and kinetics of crystallization in deeply supercooled {S}tillinger–{W}eber silicon},}\ }\href {https://doi.org/10.1063/5.0069475} {\bibfield  {journal} {\bibinfo  {journal} {J. Chem. Phys.}\ }\textbf {\bibinfo {volume} {155}},\ \bibinfo {pages} {194502} (\bibinfo {year} {2021})}\BibitemShut {NoStop}%
\bibitem [{\citenamefont {González}\ \emph {et~al.}(2014)\citenamefont {González}, \citenamefont {Menzl}, \citenamefont {Aragones}, \citenamefont {Geiger}, \citenamefont {Caupin}, \citenamefont {Abascal}, \citenamefont {Dellago},\ and\ \citenamefont {Valeriani}}]{RN2847}%
  \BibitemOpen
  \bibfield  {author} {\bibinfo {author} {\bibfnamefont {M.~A.}\ \bibnamefont {González}}, \bibinfo {author} {\bibfnamefont {G.}~\bibnamefont {Menzl}}, \bibinfo {author} {\bibfnamefont {J.~L.}\ \bibnamefont {Aragones}}, \bibinfo {author} {\bibfnamefont {P.}~\bibnamefont {Geiger}}, \bibinfo {author} {\bibfnamefont {F.}~\bibnamefont {Caupin}}, \bibinfo {author} {\bibfnamefont {J.~L.~F.}\ \bibnamefont {Abascal}}, \bibinfo {author} {\bibfnamefont {C.}~\bibnamefont {Dellago}},\ and\ \bibinfo {author} {\bibfnamefont {C.}~\bibnamefont {Valeriani}},\ }\bibfield  {title} {\enquote {\bibinfo {title} {Detecting vapour bubbles in simulations of metastable water},}\ }\href {https://doi.org/10.1063/1.4896216} {\bibfield  {journal} {\bibinfo  {journal} {The Journal of Chemical Physics}\ }\textbf {\bibinfo {volume} {141}},\ \bibinfo {pages} {18C511} (\bibinfo {year} {2014})}\BibitemShut {NoStop}%
\bibitem [{\citenamefont {Wang}, \citenamefont {Valeriani},\ and\ \citenamefont {Frenkel}(2009)}]{RN1184}%
  \BibitemOpen
  \bibfield  {author} {\bibinfo {author} {\bibfnamefont {Z.-J.}\ \bibnamefont {Wang}}, \bibinfo {author} {\bibfnamefont {C.}~\bibnamefont {Valeriani}},\ and\ \bibinfo {author} {\bibfnamefont {D.}~\bibnamefont {Frenkel}},\ }\bibfield  {title} {\enquote {\bibinfo {title} {Homogeneous bubble nucleation driven by local hot spots: {A} molecular dynamics study},}\ }\href {https://doi.org/10.1021/jp807727p} {\bibfield  {journal} {\bibinfo  {journal} {The Journal of Physical Chemistry B}\ }\textbf {\bibinfo {volume} {113}},\ \bibinfo {pages} {3776--3784} (\bibinfo {year} {2009})}\BibitemShut {NoStop}%
\bibitem [{\citenamefont {Allen}\ and\ \citenamefont {Tildesley}(1987)}]{RN154}%
  \BibitemOpen
  \bibfield  {author} {\bibinfo {author} {\bibfnamefont {M.~P.}\ \bibnamefont {Allen}}\ and\ \bibinfo {author} {\bibfnamefont {D.~J.}\ \bibnamefont {Tildesley}},\ }\href@noop {} {\emph {\bibinfo {title} {Computer Simulation of Liquids}}}\ (\bibinfo  {publisher} {Clarendon Press},\ \bibinfo {address} {Oxford},\ \bibinfo {year} {1987})\BibitemShut {NoStop}%
\bibitem [{\citenamefont {Frenkel}\ and\ \citenamefont {Smit}(2002)}]{RN51}%
  \BibitemOpen
  \bibfield  {author} {\bibinfo {author} {\bibfnamefont {D.}~\bibnamefont {Frenkel}}\ and\ \bibinfo {author} {\bibfnamefont {B.}~\bibnamefont {Smit}},\ }\href@noop {} {\emph {\bibinfo {title} {Understanding Molecular Simulation}}}\ (\bibinfo  {publisher} {Academic Press},\ \bibinfo {address} {London},\ \bibinfo {year} {2002})\BibitemShut {NoStop}%
\bibitem [{\citenamefont {Watanabe}, \citenamefont {Ito},\ and\ \citenamefont {Hu}(2012)}]{RN1330}%
  \BibitemOpen
  \bibfield  {author} {\bibinfo {author} {\bibfnamefont {H.}~\bibnamefont {Watanabe}}, \bibinfo {author} {\bibfnamefont {N.}~\bibnamefont {Ito}},\ and\ \bibinfo {author} {\bibfnamefont {C.-K.}\ \bibnamefont {Hu}},\ }\bibfield  {title} {\enquote {\bibinfo {title} {Phase diagram and universality of the {L}ennard-{J}ones gas-liquid system},}\ }\href {https://doi.org/10.1063/1.4720089} {\bibfield  {journal} {\bibinfo  {journal} {The Journal of Chemical Physics}\ }\textbf {\bibinfo {volume} {136}},\ \bibinfo {pages} {204102} (\bibinfo {year} {2012})}\BibitemShut {NoStop}%
\bibitem [{\citenamefont {Stillinger}(1963)}]{RN3144}%
  \BibitemOpen
  \bibfield  {author} {\bibinfo {author} {\bibfnamefont {J.}~\bibnamefont {Stillinger}, \bibfnamefont {Frank~H.}},\ }\bibfield  {title} {\enquote {\bibinfo {title} {Rigorous basis of the {F}renkel‐{B}and theory of association equilibrium},}\ }\href {https://doi.org/10.1063/1.1776907} {\bibfield  {journal} {\bibinfo  {journal} {The Journal of Chemical Physics}\ }\textbf {\bibinfo {volume} {38}},\ \bibinfo {pages} {1486--1494} (\bibinfo {year} {1963})}\BibitemShut {NoStop}%
\bibitem [{\citenamefont {Torrie}\ and\ \citenamefont {Valleau}(1974)}]{RN2885}%
  \BibitemOpen
  \bibfield  {author} {\bibinfo {author} {\bibfnamefont {G.~M.}\ \bibnamefont {Torrie}}\ and\ \bibinfo {author} {\bibfnamefont {J.~P.}\ \bibnamefont {Valleau}},\ }\bibfield  {title} {\enquote {\bibinfo {title} {{M}onte {C}arlo free energy estimates using non-{B}oltzmann sampling: Application to the sub-critical {L}ennard-{J}ones fluid},}\ }\href@noop {} {\bibfield  {journal} {\bibinfo  {journal} {Chem. Phys. Lett.}\ }\textbf {\bibinfo {volume} {28}},\ \bibinfo {pages} {578--581} (\bibinfo {year} {1974})}\BibitemShut {NoStop}%
\bibitem [{\citenamefont {Maibaum}(2008)}]{RN2893}%
  \BibitemOpen
  \bibfield  {author} {\bibinfo {author} {\bibfnamefont {L.}~\bibnamefont {Maibaum}},\ }\bibfield  {title} {\enquote {\bibinfo {title} {Comment on ``elucidating the mechanism of nucleation near the gas-liquid spinodal''},}\ }\href {https://doi.org/10.1103/PhysRevLett.101.019601} {\bibfield  {journal} {\bibinfo  {journal} {Physical Review Letters}\ }\textbf {\bibinfo {volume} {101}},\ \bibinfo {pages} {019601} (\bibinfo {year} {2008})}\BibitemShut {NoStop}%
\bibitem [{\citenamefont {Chakrabarty}, \citenamefont {Santra},\ and\ \citenamefont {Bagchi}(2008)}]{RN2895}%
  \BibitemOpen
  \bibfield  {author} {\bibinfo {author} {\bibfnamefont {S.}~\bibnamefont {Chakrabarty}}, \bibinfo {author} {\bibfnamefont {M.}~\bibnamefont {Santra}},\ and\ \bibinfo {author} {\bibfnamefont {B.}~\bibnamefont {Bagchi}},\ }\bibfield  {title} {\enquote {\bibinfo {title} {{C}hakrabarty, {S}antra, and {B}agchi {R}eply},}\ }\href {https://doi.org/10.1103/PhysRevLett.101.019602} {\bibfield  {journal} {\bibinfo  {journal} {Physical Review Letters}\ }\textbf {\bibinfo {volume} {101}},\ \bibinfo {pages} {019602} (\bibinfo {year} {2008})}\BibitemShut {NoStop}%
\bibitem [{\citenamefont {Leyssale}, \citenamefont {Delhommelle},\ and\ \citenamefont {Millot}(lack)}]{RN2936}%
  \BibitemOpen
  \bibfield  {author} {\bibinfo {author} {\bibfnamefont {J.-M.}\ \bibnamefont {Leyssale}}, \bibinfo {author} {\bibfnamefont {J.}~\bibnamefont {Delhommelle}},\ and\ \bibinfo {author} {\bibfnamefont {C.}~\bibnamefont {Millot}},\ }\bibfield  {title} {\enquote {\bibinfo {title} {Atomistic simulation of the homogeneous nucleation and of the growth of {N}$_2$ crystallites},}\ }\href {https://doi.org/10.1063/1.1862626} {\bibfield  {journal} {\bibinfo  {journal} {The Journal of Chemical Physics}\ }\textbf {\bibinfo {volume} {122}},\ \bibinfo {pages} {104510} (\bibinfo {year} {2005\color{black}})}\BibitemShut {NoStop}%
\bibitem [{\citenamefont {ten Wolde}\ and\ \citenamefont {Frenkel}(1997)}]{RN2909}%
  \BibitemOpen
  \bibfield  {author} {\bibinfo {author} {\bibfnamefont {P.~R.}\ \bibnamefont {ten Wolde}}\ and\ \bibinfo {author} {\bibfnamefont {D.}~\bibnamefont {Frenkel}},\ }\bibfield  {title} {\enquote {\bibinfo {title} {Enhancement of protein crystal nucleation by critical density fluctuations},}\ }\href {https://doi.org/10.1126/science.277.5334.1975} {\bibfield  {journal} {\bibinfo  {journal} {Science}\ }\textbf {\bibinfo {volume} {277}},\ \bibinfo {pages} {1975--1978} (\bibinfo {year} {1997})}\BibitemShut {NoStop}%
\bibitem [{\citenamefont {Puibasset}(2022)}]{RN2954}%
  \BibitemOpen
  \bibfield  {author} {\bibinfo {author} {\bibfnamefont {J.}~\bibnamefont {Puibasset}},\ }\bibfield  {title} {\enquote {\bibinfo {title} {A general relation between the largest nucleus and all nuclei distributions for free energy calculations},}\ }\href {https://doi.org/10.1063/5.0121580} {\bibfield  {journal} {\bibinfo  {journal} {The Journal of Chemical Physics}\ }\textbf {\bibinfo {volume} {157}},\ \bibinfo {pages} {191102} (\bibinfo {year} {2022})}\BibitemShut {NoStop}%
\bibitem [{\citenamefont {Porion}\ and\ \citenamefont {Puibasset}(2024)}]{RN3149}%
  \BibitemOpen
  \bibfield  {author} {\bibinfo {author} {\bibfnamefont {P.}~\bibnamefont {Porion}}\ and\ \bibinfo {author} {\bibfnamefont {J.}~\bibnamefont {Puibasset}},\ }\bibfield  {title} {\enquote {\bibinfo {title} {A statistical analysis of the first stages of freezing and melting of {L}ennard-{J}ones particles: {N}umber and size distributions of transient nuclei},}\ }\href {https://doi.org/10.1063/5.0216704} {\bibfield  {journal} {\bibinfo  {journal} {The Journal of Chemical Physics}\ }\textbf {\bibinfo {volume} {161}},\ \bibinfo {pages} {074501} (\bibinfo {year} {2024})}\BibitemShut {NoStop}%
\bibitem [{\citenamefont {Stukowski}(2010)}]{RN3044}%
  \BibitemOpen
  \bibfield  {author} {\bibinfo {author} {\bibfnamefont {A.}~\bibnamefont {Stukowski}},\ }\bibfield  {title} {\enquote {\bibinfo {title} {Visualization and analysis of atomistic simulation data with {OVITO}–the {O}pen {V}isualization {T}ool},}\ }\href {https://doi.org/10.1088/0965-0393/18/1/015012} {\bibfield  {journal} {\bibinfo  {journal} {Modelling and Simulation in Materials Science and Engineering}\ }\textbf {\bibinfo {volume} {18}},\ \bibinfo {pages} {015012} (\bibinfo {year} {2010})}\BibitemShut {NoStop}%
\bibitem [{\citenamefont {Sharma}\ and\ \citenamefont {Escobedo}(2018)}]{RN3194}%
  \BibitemOpen
  \bibfield  {author} {\bibinfo {author} {\bibfnamefont {A.~K.}\ \bibnamefont {Sharma}}\ and\ \bibinfo {author} {\bibfnamefont {F.~A.}\ \bibnamefont {Escobedo}},\ }\bibfield  {title} {\enquote {\bibinfo {title} {Nucleus-size pinning for determination of nucleation free-energy barriers and nucleus geometry},}\ }\href {https://doi.org/10.1063/1.5021602} {\bibfield  {journal} {\bibinfo  {journal} {The Journal of Chemical Physics}\ }\textbf {\bibinfo {volume} {148}},\ \bibinfo {pages} {184104} (\bibinfo {year} {2018})}\BibitemShut {NoStop}%
\bibitem [{\citenamefont {Prestipino}, \citenamefont {Laio},\ and\ \citenamefont {Tosatti}(2012)}]{RN3197}%
  \BibitemOpen
  \bibfield  {author} {\bibinfo {author} {\bibfnamefont {S.}~\bibnamefont {Prestipino}}, \bibinfo {author} {\bibfnamefont {A.}~\bibnamefont {Laio}},\ and\ \bibinfo {author} {\bibfnamefont {E.}~\bibnamefont {Tosatti}},\ }\bibfield  {title} {\enquote {\bibinfo {title} {Systematic improvement of classical nucleation theory},}\ }\href {https://doi.org/10.1103/PhysRevLett.108.225701} {\bibfield  {journal} {\bibinfo  {journal} {Physical Review Letters}\ }\textbf {\bibinfo {volume} {108}},\ \bibinfo {pages} {225701} (\bibinfo {year} {2012})}\BibitemShut {NoStop}%
\bibitem [{\citenamefont {Aasen}\ \emph {et~al.}(2023)\citenamefont {Aasen}, \citenamefont {Wilhelmsen}, \citenamefont {Hammer},\ and\ \citenamefont {Reguera}}]{RN3147}%
  \BibitemOpen
  \bibfield  {author} {\bibinfo {author} {\bibfnamefont {A.}~\bibnamefont {Aasen}}, \bibinfo {author} {\bibfnamefont {O.}~\bibnamefont {Wilhelmsen}}, \bibinfo {author} {\bibfnamefont {M.}~\bibnamefont {Hammer}},\ and\ \bibinfo {author} {\bibfnamefont {D.}~\bibnamefont {Reguera}},\ }\bibfield  {title} {\enquote {\bibinfo {title} {Free energy of critical droplets—from the binodal to the spinodal},}\ }\href {https://doi.org/10.1063/5.0142533} {\bibfield  {journal} {\bibinfo  {journal} {The Journal of Chemical Physics}\ }\textbf {\bibinfo {volume} {158}},\ \bibinfo {pages} {114108} (\bibinfo {year} {2023})}\BibitemShut {NoStop}%
\bibitem [{\citenamefont {Gispen}\ \emph {et~al.}(2024)\citenamefont {Gispen}, \citenamefont {Espinosa}, \citenamefont {Sanz}, \citenamefont {Vega},\ and\ \citenamefont {Dijkstra}}]{RN3183}%
  \BibitemOpen
  \bibfield  {author} {\bibinfo {author} {\bibfnamefont {W.}~\bibnamefont {Gispen}}, \bibinfo {author} {\bibfnamefont {J.~R.}\ \bibnamefont {Espinosa}}, \bibinfo {author} {\bibfnamefont {E.}~\bibnamefont {Sanz}}, \bibinfo {author} {\bibfnamefont {C.}~\bibnamefont {Vega}},\ and\ \bibinfo {author} {\bibfnamefont {M.}~\bibnamefont {Dijkstra}},\ }\bibfield  {title} {\enquote {\bibinfo {title} {Variational umbrella seeding for calculating nucleation barriers},}\ }\href {https://doi.org/10.1063/5.0204540} {\bibfield  {journal} {\bibinfo  {journal} {The Journal of Chemical Physics}\ }\textbf {\bibinfo {volume} {160}},\ \bibinfo {pages} {174501} (\bibinfo {year} {2024})}\BibitemShut {NoStop}%
\bibitem [{\citenamefont {Tolman}(1949)}]{RN629}%
  \BibitemOpen
  \bibfield  {author} {\bibinfo {author} {\bibfnamefont {R.~C.}\ \bibnamefont {Tolman}},\ }\bibfield  {title} {\enquote {\bibinfo {title} {The effect of droplet size on surface tension},}\ }\href@noop {} {\bibfield  {journal} {\bibinfo  {journal} {The Journal of Chemical Physics}\ }\textbf {\bibinfo {volume} {17}},\ \bibinfo {pages} {333--337} (\bibinfo {year} {1949})}\BibitemShut {NoStop}%
\bibitem [{\citenamefont {Sampayo}\ \emph {et~al.}(2010)\citenamefont {Sampayo}, \citenamefont {Malijevský}, \citenamefont {Müller}, \citenamefont {de~Miguel},\ and\ \citenamefont {Jackson}}]{RN1293}%
  \BibitemOpen
  \bibfield  {author} {\bibinfo {author} {\bibfnamefont {J.~G.}\ \bibnamefont {Sampayo}}, \bibinfo {author} {\bibfnamefont {A.}~\bibnamefont {Malijevský}}, \bibinfo {author} {\bibfnamefont {E.~A.}\ \bibnamefont {Müller}}, \bibinfo {author} {\bibfnamefont {E.}~\bibnamefont {de~Miguel}},\ and\ \bibinfo {author} {\bibfnamefont {G.}~\bibnamefont {Jackson}},\ }\bibfield  {title} {\enquote {\bibinfo {title} {Communications: Evidence for the role of fluctuations in the thermodynamics of nanoscale drops and the implications in computations of the surface tension},}\ }\href@noop {} {\bibfield  {journal} {\bibinfo  {journal} {The Journal of Chemical Physics}\ }\textbf {\bibinfo {volume} {132}},\ \bibinfo {pages} {141101} (\bibinfo {year} {2010})}\BibitemShut {NoStop}%
\bibitem [{\citenamefont {Rehner}\ and\ \citenamefont {Gross}(2018)}]{RN3154}%
  \BibitemOpen
  \bibfield  {author} {\bibinfo {author} {\bibfnamefont {P.}~\bibnamefont {Rehner}}\ and\ \bibinfo {author} {\bibfnamefont {J.}~\bibnamefont {Gross}},\ }\bibfield  {title} {\enquote {\bibinfo {title} {Surface tension of droplets and tolman lengths of real substances and mixtures from density functional theory},}\ }\href {https://doi.org/10.1063/1.5020421} {\bibfield  {journal} {\bibinfo  {journal} {The Journal of Chemical Physics}\ }\textbf {\bibinfo {volume} {148}},\ \bibinfo {pages} {164703} (\bibinfo {year} {2018})}\BibitemShut {NoStop}%
\bibitem [{\citenamefont {Julin}\ \emph {et~al.}(2010)\citenamefont {Julin}, \citenamefont {Napari}, \citenamefont {Merikanto},\ and\ \citenamefont {Vehkamäki}}]{RN3182}%
  \BibitemOpen
  \bibfield  {author} {\bibinfo {author} {\bibfnamefont {J.}~\bibnamefont {Julin}}, \bibinfo {author} {\bibfnamefont {I.}~\bibnamefont {Napari}}, \bibinfo {author} {\bibfnamefont {J.}~\bibnamefont {Merikanto}},\ and\ \bibinfo {author} {\bibfnamefont {H.}~\bibnamefont {Vehkamäki}},\ }\bibfield  {title} {\enquote {\bibinfo {title} {A thermodynamically consistent determination of surface tension of small {L}ennard-{J}ones clusters from simulation and theory},}\ }\href {https://doi.org/10.1063/1.3456184} {\bibfield  {journal} {\bibinfo  {journal} {The Journal of Chemical Physics}\ }\textbf {\bibinfo {volume} {133}},\ \bibinfo {pages} {044704} (\bibinfo {year} {2010})}\BibitemShut {NoStop}%
\end{thebibliography}%

\end{document}